\documentclass[prd,twocolumn,aps,superscriptaddress,nofootinbib]{revtex4-2}
\usepackage[utf8]{inputenc}

\usepackage{multirow}
\usepackage{graphicx}
\usepackage{bm}
\usepackage{wrapfig}
\usepackage{mathtools}
\usepackage{amssymb}
\usepackage[svgnames,table]{xcolor}
\usepackage{subfig}
\usepackage{comment}

\usepackage{microtype}
\usepackage{fouriernc}
\usepackage{natbib}

\usepackage[colorlinks=true, citecolor=DimGray,
linkcolor=SlateBlue, urlcolor=Navy]{hyperref}

\usepackage[normalem]{ulem}

\usepackage{caption}
\usepackage{cleveref}
\crefformat{section}{\S#2#1#3}
\crefmultiformat{section}{\S\S#2#1#3}{and~#2#1#3}{, #2#1#3}{, and~#2#1#3}

\begin{document}

\title{Optimizing Gravitational-Wave Detector Design for Squeezed Light}

\author{Jonathan W. Richardson}
\email{jonathan.richardson@ucr.edu}
\affiliation{Department of Physics and Astronomy, University of California, Riverside, Riverside, CA 92521, USA}

\author{Swadha Pandey}
\affiliation{Department of Physics, Indian Institute of Technology Kanpur, Kanpur, UP 208016, India}

\author{Edita Bytyqi}
\affiliation{Department of Applied Physics and Applied Mathematics, Columbia University, New York, NY 10027, USA}

\author{Tega Edo}
\affiliation{LIGO Laboratory, California Institute of Technology, Pasadena, CA 91125, USA}

\author{Rana X. Adhikari}
\affiliation{LIGO Laboratory, California Institute of Technology, Pasadena, CA 91125, USA}

\date{\today}

\begin{abstract}
Achieving the quantum noise targets of third-generation detectors will require 10~dB of squeezed-light enhancement as well as megawatt laser power in the interferometer arms---both of which require unprecedented control of the internal optical losses. In this work, we present a novel optimization approach to gravitational-wave detector design aimed at maximizing the robustness to common, yet unavoidable, optical fabrication and installation errors, which have caused significant loss in Advanced LIGO. As a proof of concept, we employ these techniques to perform a two-part optimization of the LIGO A+ design. First, we optimize the arm cavities for reduced scattering loss in the presence of point absorbers, as currently limit the operating power of Advanced LIGO. Then, we optimize the signal recycling cavity for maximum squeezing performance, accounting for realistic errors in the positions and radii of curvature of the optics. Our findings suggest that these techniques can be leveraged to achieve substantially greater quantum noise performance in current and future gravitational-wave detectors.
\end{abstract}

\maketitle

\section{Introduction}
\label{sec:intro}
In the last six years, Advanced LIGO and Virgo have established gravitational waves as a new observational probe of the Universe. With projected improvements in gravitational-wave detector sensitivities, new tests of gravity, cosmology, and dense nuclear matter will become possible within the next decade. Higher sensitivity in the 200~Hz--1~kHz band will resolve the ringdown radiation of newly coalesced black holes, detecting or constraining potential quantum modifications at the event horizon~\cite{Skenderis:2008, Cardoso:2017, Brustein:2018}. Higher sensitivity in the 1.5--5~kHz band will resolve binary neutron star mergers to the moment of coalescence, illuminating the neutron star equation of state~\cite{Markakis:2009, Read:2013}. More frequent detections of binary neutron star mergers will also enable independent measurement of the Hubble constant to high precision~\cite{Chen:2018}, addressing the growing tension between cosmic microwave background and local distance ladder measurements~\cite{Riess:2019}.

The LIGO detectors are sensitive to gravitational waves in a broad frequency band ranging from 20~Hz to 5~kHz. Across this band, the limiting source of instrumental noise transitions from sensing and controls noise, below roughly 30~Hz, to Brownian noise of the dielectric optical coatings, up to roughly 200~Hz, and finally to quantum noise at higher frequencies (see \cite{Buikema:2020} for a thorough review of the LIGO noise budget). In laser interferometers, quantum noise arises not from the positional uncertainties of the mirrors, but from the quantization of the electromagnetic field used to interrogate their positions~\cite{Caves:1980, Caves:1981}. This effect, commonly described as ``shot noise,'' arises from ground-state fluctuations of the vacuum field, which enter the interferometer and beat with the circulating laser field. The interference of the two fields produces intensity fluctuations which modulate the interferometer output signal. These fluctuations also apply force to the mirrors via radiation pressure, producing actual mirror displacements at low frequencies. Shot noise can be reduced through two means: higher laser power in the interferometer, which increases number of photons incident on the beamsplitter, and the injection of squeezed quantum states of light. Both are critical to improving the high-frequency sensitivity of gravitational-wave detectors.

In the third observing run, the Advanced LIGO detectors operated with roughly 250~kW of resonating power inside the arm cavities~\cite{Buikema:2020}---still only one third of their 750~kW design power. Recent tests in both detectors have shown that as the injected laser power is increased, the arm cavity optical gain severely decays due to increasing internal loss~\cite{Buikema:2020}. The source of this loss has been identified as sub-millimeter, highly absorbing defects in the optical coatings known as \emph{point absorbers}. In situ wavefront sensors have detected their presence on at least four of the eight currently installed test masses~\cite{Buikema:2020, Brooks:2021}. Point absorbers appear to originate during the coating deposition process, although it is still not understood how these contaminants enter the coating nor to what extent they can be eliminated. Each point absorber absorbs roughly 80~ppb of the total incident power, or 20~mW when exposed to 250~kW. The extremely localized heating induces a sharply peaked thermoelastic deformation of the mirror surface, which scatters power into higher-order spatial modes~\cite{Brooks:2021}. To achieve higher operating power, point absorber losses must be mitigated.

Beginning also in the third observing run, squeezed light was injected into both Advanced LIGO detectors~\cite{Tse:2019}. Squeezed light allows for the engineering of the electromagnetic vacuum state that enters the interferometer. Quantum fluctuations of the vacuum field, initially distributed uniformly between the amplitude and phase quadratures, are redistributed so that they are suppressed in the phase quadrature, containing the gravitational-wave signal, and amplified in the unsensed amplitude quadrature. In Advanced LIGO, squeezed vacuum field is generated via degenerate optical parametric amplification~\cite{Ciriolo:2017} and injected into the interferometer output port. During the third observing run, a shot noise reduction factor of roughly 3~dB was achieved in each detector~\cite{Tse:2019}.

Although the injected level of squeezing can be high, the observed level of squeezing at the interferometer output depends on the amount of entanglement remaining in the squeezed field. Losses within the detector lead directly to decoherence of the squeezed state, limiting the quantum noise reduction~\cite{Miao:2019}. Losses arise from scattering, imperfect transmissivity or reflectivity of optics, photodetector quantum efficiency, and spatial mode-mismatch between the optical cavities. For the interferometer, the largest source of loss is mode-mismatch between the coupled laser cavities. For example, in the Advanced LIGO detectors, the mode-matching loss between the arm cavities and the output mode cleaner cavity alone is measured to be 10\%. It has been demonstrated that this loss can be attributed to practical, and largely irreducible, limitations in the fabrication and hand-positioning of the interferometer optics~\cite{Perreca:2020}. For third-generation detectors~\cite{Reitze:2019, Abernathy:2011}, reducing the internal mode-matching losses to $\sim 1\%$ levels is imperative.

In this work, we present a novel optimization approach to gravitational-wave detector design. It is aimed at maximizing the robustness to common optical fabrication and installation errors, which introduce losses that degrade the optical gain and squeezing performance. Under this approach, design performance is assessed and improved statistically, over thousands of trials in which realistic random errors are assumed in the surface figures and positions of the optics. As a proof of concept, we employ these techniques to perform a two-part optimization of the LIGO A+ interferometers, planned to become operational in 2025~\cite{Fritschel:2020b}. First, in \S\ref{sec:arm} we modify the arm cavities for reduced scattering loss in the presence of point absorbers. Then, in \S\ref{sec:src} we optimize the signal recycling cavity (SRC) for maximum squeezing performance, accounting for realistic errors in the positions and radii of curvature of the optics. Our findings suggest that these techniques can be leveraged to achieve substantially greater quantum noise performance in current and future gravitational-wave detectors. Finally, in \S\ref{sec:conclusions} we summarize and discuss future extensions of this work.

\section{Arm cavity design}
\label{sec:arm}

\begin{figure*}[ht]
    \centering
    \subfloat{\includegraphics[width=0.48\textwidth]{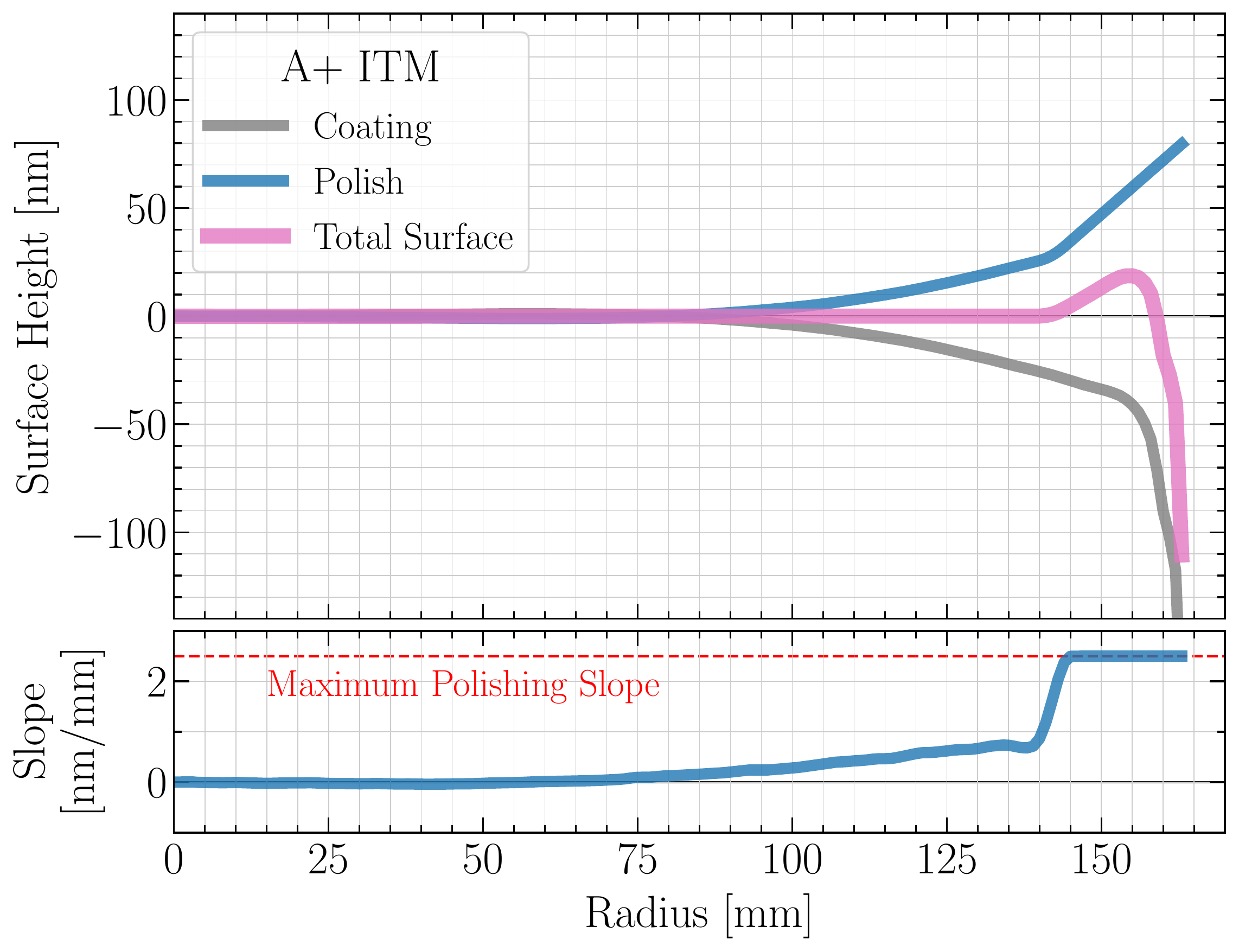}}
    \hfill
    \subfloat{\includegraphics[width=0.48\textwidth]{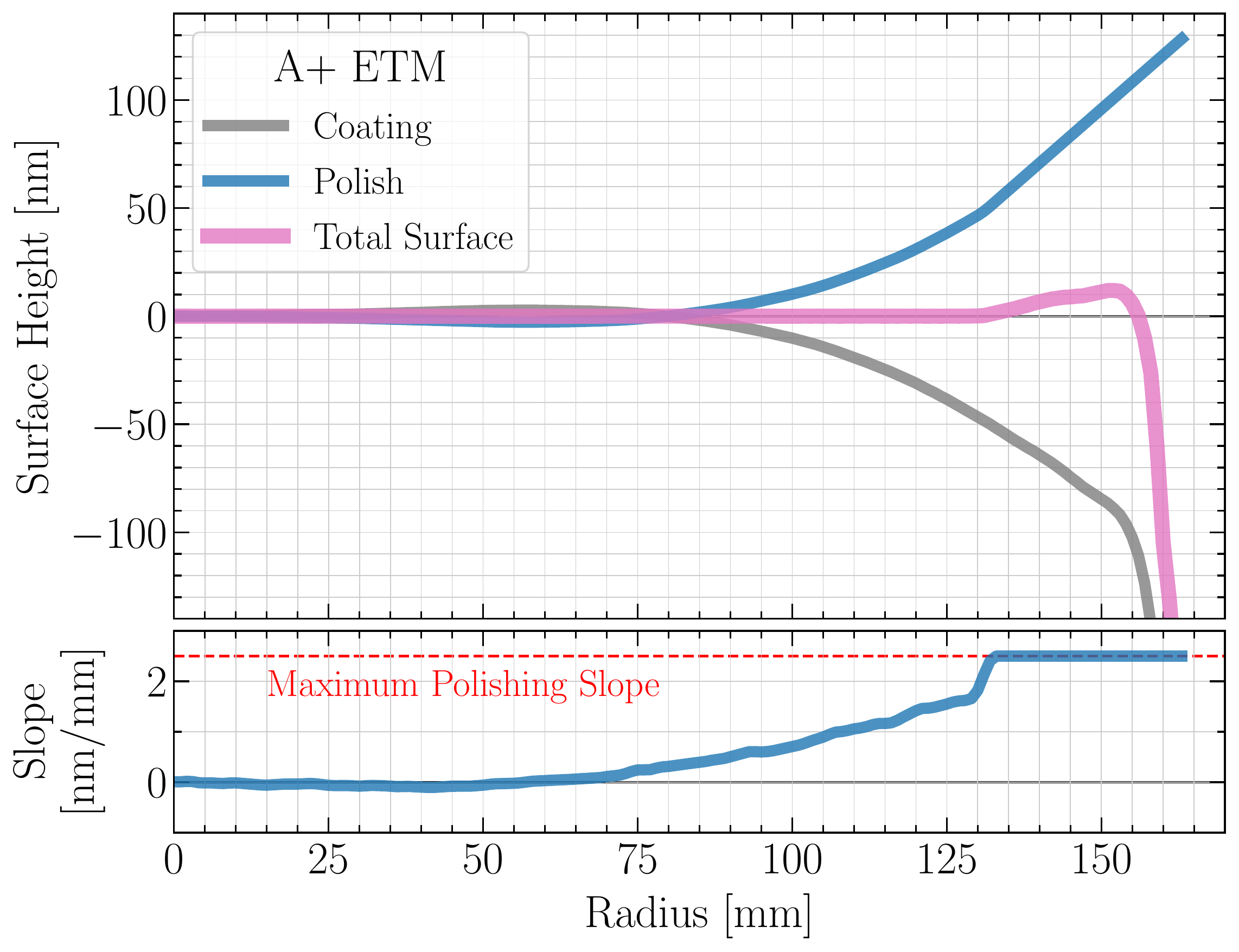}}
    \caption{Proposed surface profiles for the LIGO A+ input test masses (ITM; left) and end test masses (ETM; right). In each panel, the total surface figure (pink curve) is the sum of the polishing profile (blue curve) and the optical coating nonuniformity (grey curve). Based on current fabrication capabilities, the polishing slope is restricted to $\le 2.5$~nm/mm, as shown in the lower panels.}
    \label{fig:profiles}
\end{figure*}

In the Advanced LIGO arm cavities, point absorbers on the mirror surfaces disproportionately scatter power into 7th-order spatial modes. Although a point-absorber-induced deformation scatters power into many higher-order modes (HOM), the Fabry–Perot cavity resonantly {\em enhances or suppresses} each mode as a function of the roundtrip phase it accumulates in the cavity. This effect was first analyzed for static deformations by Vajente~\cite{Vajente:2014} and extended to power-dependent surface deformations from point absorbers by Brooks~{\em et al.}~\cite{Brooks:2021}, who showed that the power loss from the fundamental mode to the $mn$-th HOM is approximately
\begin{equation}
    \mathcal{L}_{mn} = a_{00|mn}^2\, g_{mn} \;.
    \label{eqn:loss_by_mode}
\end{equation}
The first term, $a_{00|mn}$, is the single-bounce amplitude scattering from the fundamental mode to the $mn$-th HOM when reflected off the deformed mirror surface. The second term, $g_{mn}$, is the optical gain of that HOM, which depends on the cavity geometry and the actual (nonideal) surface profiles of the two mirrors:
\begin{equation}
    g_{mn} = \frac{1 - r_{1}^{\prime \,2} \,r_{2}^{\prime \,2}}{1 + r_{1}^{\prime \,2}\,r_{2}^{\prime \,2} } \, \frac{1}{1 - \frac{2\, r_{1}^{\prime} \, r_{2}^{\prime} }{1 + r_{1}^{\prime\,2} \, r_{2}^{\prime\,2}} \,\cos \left[ \Phi_{mn}\right]} \;.
    \label{eqn:optical_gain_factor}
\end{equation}
The factors $r_{1}^{\prime}$ and $r_{2}^{\prime}$ are the effective amplitude reflectivities of the input test mass (ITM) and the end test mass (ETM), respectively, accounting for mode-dependent clipping losses, and $\Phi_{mn}$ is the additional roundtrip phase that the HOM accumulates relative to the fundamental mode. In the LIGO arm cavities, modes of order~7, by coincidence, are nearly co-resonant with the fundamental mode, leading to optical gain factors $g_{mn}$ up to 100~times larger than those for non-resonant modes.

Thus, to reduce point absorber losses and achieve higher operational power in LIGO A+, our design objective is to fully eliminate mode co-resonances below order~8 in the arm cavities. In principle, this could be achieved by adjusting the arm cavity parameters (the arm length and the radii of curvature of the test masses) for a more favorable transverse mode spacing. However, a significant change of the cavity parameters is precluded by other operational constraints. The 4~km arm length is constrained by the existing infrastructure to approximately $\pm 2$~m of the current length. With only the radii of curvature of the two mirrors free to vary, it is not possible to maintain the current beam sizes on both optics. A smaller beam size results in increased coating Brownian noise---unacceptable for the A+ design, which is already thermal-noise-limited across its mid-frequency band~\cite{Fritschel:2020b}. A larger beam size, on the other hand, results in unacceptably higher clipping losses inside the arms and the signal recycling cavity. Thus, the problem is overconstrained from the perspective of a standard cavity design approach using spherical optics.

In this section, we demonstrate that by applying novel, nonspherical surface profiles to the LIGO test masses, the mode~7 co-resonances can be eliminated \emph{without} incurring any increase in coating thermal noise or clipping loss. Our approach exploits the large difference in transverse spatial confinement between the fundamental mode and 7th-order modes. Each mirror profile is spherical in the central region, where fundamental mode power is concentrated, but assumes a sharply nonspherical shape at the outermost radii, where the incident power is almost purely in higher-order modes. We show that the outer surface profile can be tailored to control the roundtrip phase $\Phi_{mn}$ (see Eq.~\ref{eqn:optical_gain_factor}) that an HOM accumulates relative to the fundamental mode. This provides a means to suppress problematic higher-order modes while {\em negligibly altering} the fundamental cavity mode, leading to a significant loss reduction in the presence of scattering sources such as point absorbers.

\subsection{Nonspherical test mass profiles}
\label{sec:nonspherical_profiles}

For mirror fabrication in the A+ era, LIGO has the ability to specify an arbitrary (nonspherical) polishing figure. Internal discussions with optics manufacturers have indicated that an arbitrary radial profile, subject to a maximum slope of 2.5~nm/mm, could be produced with high confidence, with a possibility that an even steeper polishing slope could be achieved. Fig.~\ref{fig:profiles} shows our proposed surface profiles for the LIGO input test masses (ITM) and end test masses (ETM). In each panel, the polishing figure (blue curve) both compensates the expected nonuniformity of the optical coating (grey curve) and adds a nonspherical edge component to produce the total surface figure (pink curve). To remain within demonstrated fabrication limits, we restrict the polishing slope to 2.5~nm/mm, as shown in the lower panels of Fig.~\ref{fig:profiles}. In A+, a new coating material with improved thermal noise performance, $\rm Ti O_2$-doped $\rm Ge O_2$~\cite{Vajente:2021}, is expected to replace the $\rm Ti O_2$-doped $\rm Ta_2 O_5$ coatings used in Advanced LIGO~\cite{Amato:2019, Granata:2020}. Accordingly, we estimate the A+ coating nonuniformity as the measured nonuniformity of the LIGO O4 coating plume, which will be reused to produce the A+ optics, multiplied by the relative coating thickness required to achieve the same reflectivity with the new material (1.2 for the ITM and 1.5 for the ETM).

\begin{figure}[t]
    \centering
    \includegraphics[width=0.48\textwidth]{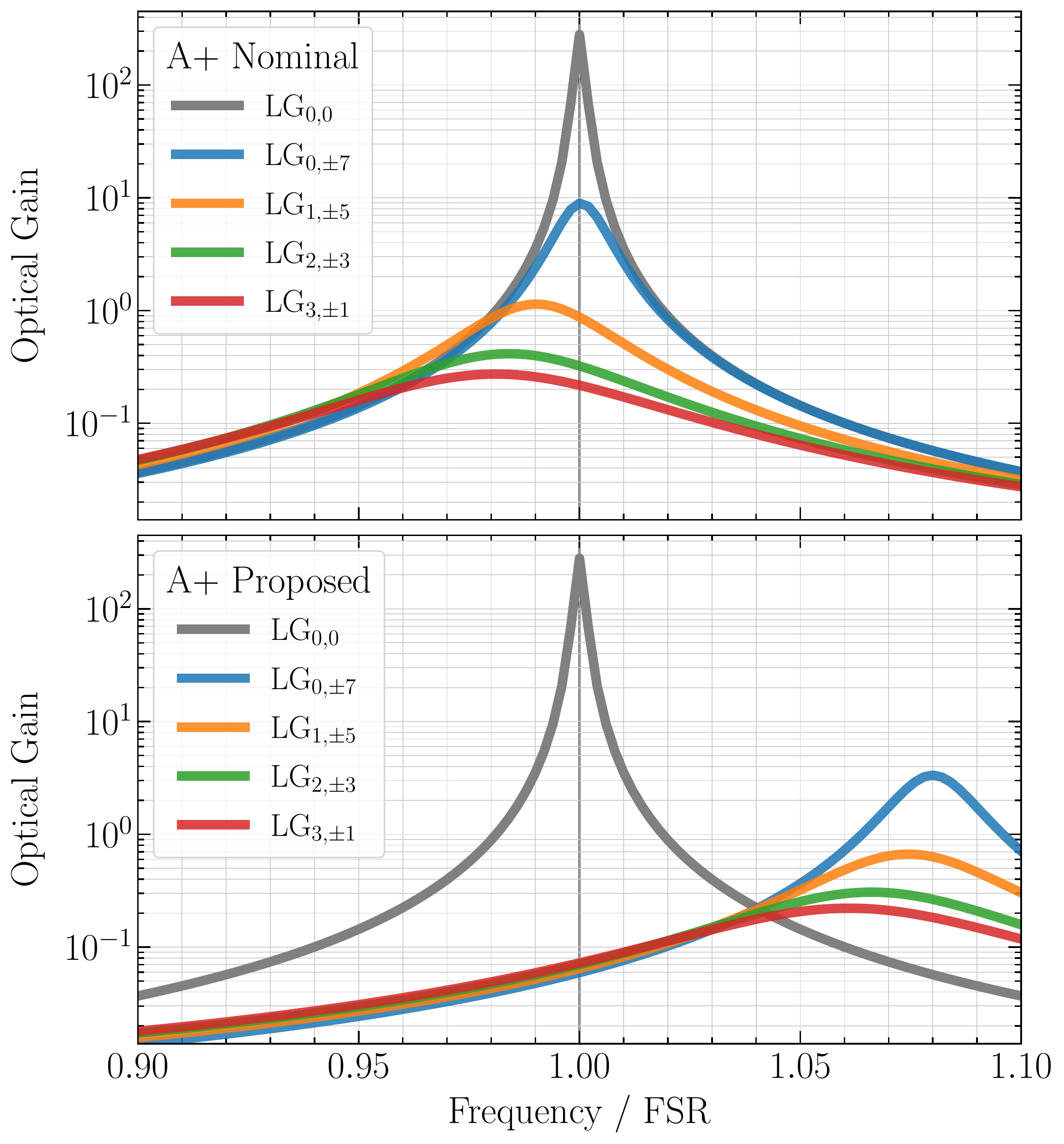}
    \caption{Optical gain $g_{mn}$ of the 7th-order Laguerre-Gauss (LG) modes in the LIGO A+ arm cavities, as a function of frequency detuning from the fundamental mode resonance. \emph{Top:} The nominal resonance locations with a spherical test mass polish. \emph{Bottom:} The new resonance locations after including the compensation polish shown in Fig.~\ref{fig:profiles}~(blue curves). In both panels, coating absorption of 120~mW per test mass is assumed.}
    \label{fig:arm_scan}
\end{figure}

Fig.~\ref{fig:arm_scan} illustrates how these profiles eliminate the arm cavity modal degeneracy. First, for comparison, the top panel shows the nominal locations of the 7th-order Laguerre-Gauss (LG) mode resonances, assuming a spherical mirror polish. With higher cavity power (or coating absorptivity), the resonances shift toward higher frequency due to the increasing residual thermal deformation of the test masses. Although ring heaters compensate the central heating due to uniform coating absorption, the ring heaters ``overcorrect'' the mirror surface at large radii, resulting in a net profile that steeply rises near the edge of the optic~\cite{Brooks:2016}. Here, we assume 120~mW of coating absorption per test mass, corresponding to a cavity power of 400~kW for absorptivity at the level of the Advanced LIGO coatings. The bottom panel shows the new locations of the 7th-order mode resonances after including the compensation polish shown in Fig.~\ref{fig:profiles}~(blue curves). The polishing profiles are designed to shift the 7th-order mode resonances rightward, toward higher frequency, where any degree of thermal distortion now strictly shifts them \emph{further away} from co-resonance. This has significant implications for the loss performance of the arm cavities, as discussed in the following section.

\subsection{Loss performance improvement}
\label{sec:arm_loss_improvement}
We now assess the impact of our proposed compensation polish on the arm cavity loss. For this, we consider two scenarios, {\it with} (``proposed'') and {\it without} (``nominal'') our proposed modification. The ``nominal'' test mass profiles (with spherical power subtracted) are equal to the grey curves in Fig.~\ref{fig:profiles}. The ``proposed'' test mass profiles (again with spherical power subtracted) are equal to the pink curves in Fig.~\ref{fig:profiles}. For each set of profiles, we perform numerical simulations of an A+ arm cavity using \textsc{SIS}~\cite{Yamamoto:2008}, an FFT-based optical simulation package. The model includes all thermoelastic effects: (1) uniform coating absorption and (2) optimal ring heater compensation, to maintain constant mode-matching of the arm to the recycling cavities. Throughout, we assume coating absorption at the average level of the Advanced LIGO test masses, 0.3~ppm. We also assume a fixed high-angle scattering loss of 25~ppm per optic. To account for realistic nonidealities, which could unequally impact the two designs, all loss analyses are performed as Monte Carlo simulations over 1000~trials with random beam miscenterings and surface roughnesses. Beam miscenterings on the ITM and ETM are independently drawn from a Gaussian distribution with zero mean and a standard deviation of 5~mm. Surface roughness profiles are randomly generated by \textsc{SIS} with a power spectral density chosen to match that of the current Advanced LIGO optics~\cite{Pinard:2017}.

\begin{figure}[t]
    \centering
    \includegraphics[width=0.48\textwidth]{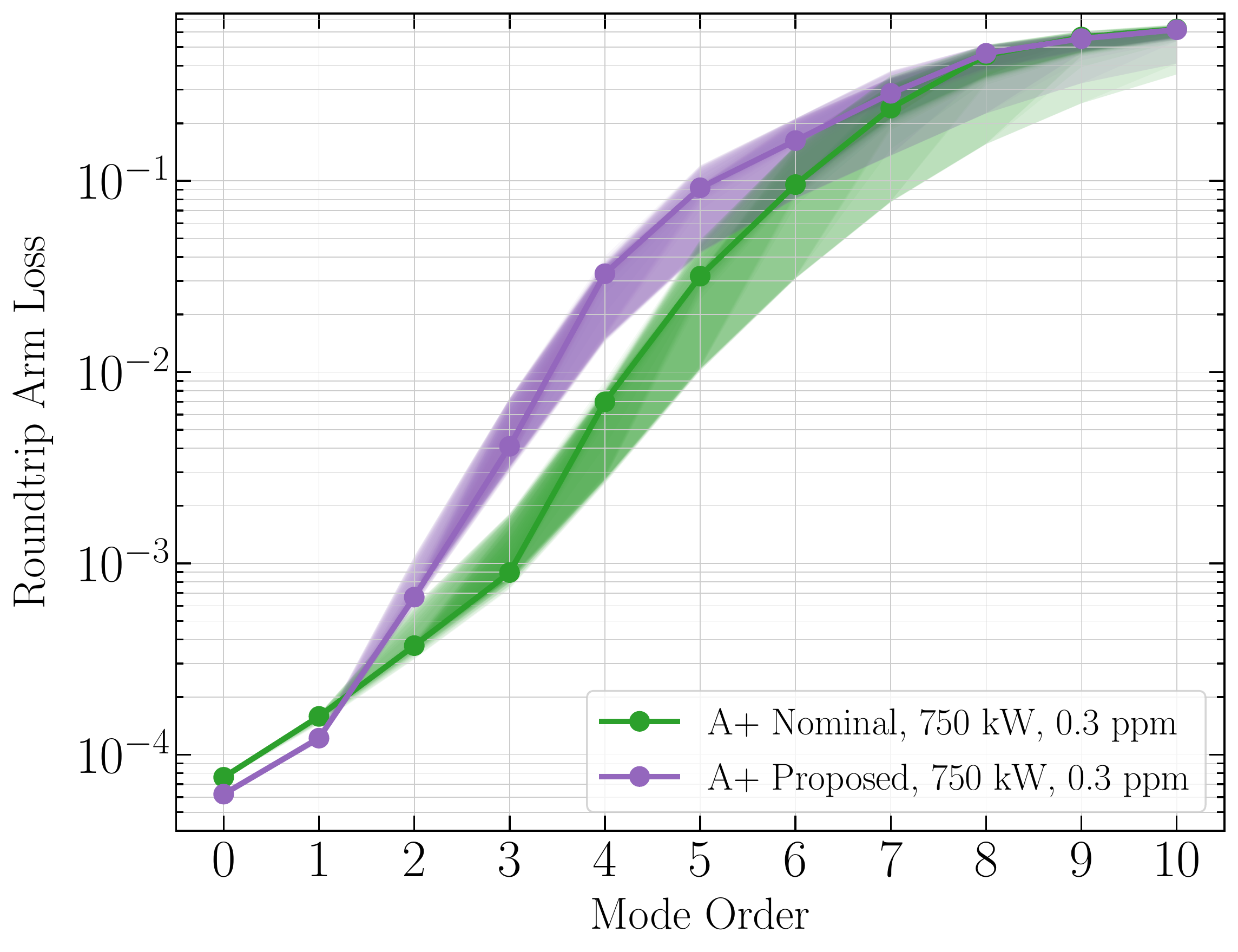}
    \caption{Arm cavity loss as a function of mode order. Each curve represents the median loss in 1000 trials with random miscenterings and surface roughnesses, averaged over all Laguerre-Gauss modes ${\rm LG}_{p,l}$ of order $N=2p + |l|$. The shading represents the 16th and 84th percentiles of the loss distributions over all trials and modes. The proposed mirror profiles achieve significantly enhanced higher-order mode dissipation, with no increase in fundamental mode loss.}
    \label{fig:loss_v_mode_order}
\end{figure}

First, we evaluate the baseline loss performance of both designs in the absence of scattering sources. The aim of our compensated design is to achieve the HOM frequency shifts outlined in \S\ref{sec:nonspherical_profiles} without worsening the roundtrip loss of the fundamental mode. Fig.~\ref{fig:loss_v_mode_order} shows the roundtrip arm loss under each design as a function of mode order, at an arm power of 750~kW. The curves represent the median loss values over all randomized trials, and averaged over all modes ${\rm LG}_{p,l}$ of order $N=2p + |l|$. The shading represents the 16th and 84th percentiles of the loss distributions across all trials and modes. Our results indicate that the proposed test mass profiles do not increase loss in the fundamental mode, but they do significantly increase the losses of HOMs above order~2. While the design objective in \S\ref{sec:nonspherical_profiles} was only to shift resonance frequencies of HOMs, the larger dissipation of HOMs is an added advantage that helps to further reduce their optical gain. Enhancing the dissipation of certain HOMs may be also relevant for improving the damping of parametric instabilities in gravitational-wave detectors~\cite{Evans:2015}. For this reason, we include a full breakdown of the dissipation per optical mode in Appendix~\ref{sec:loss_by_mode}.

\begin{figure}[t]
    \centering
    \includegraphics[width=0.48\textwidth]{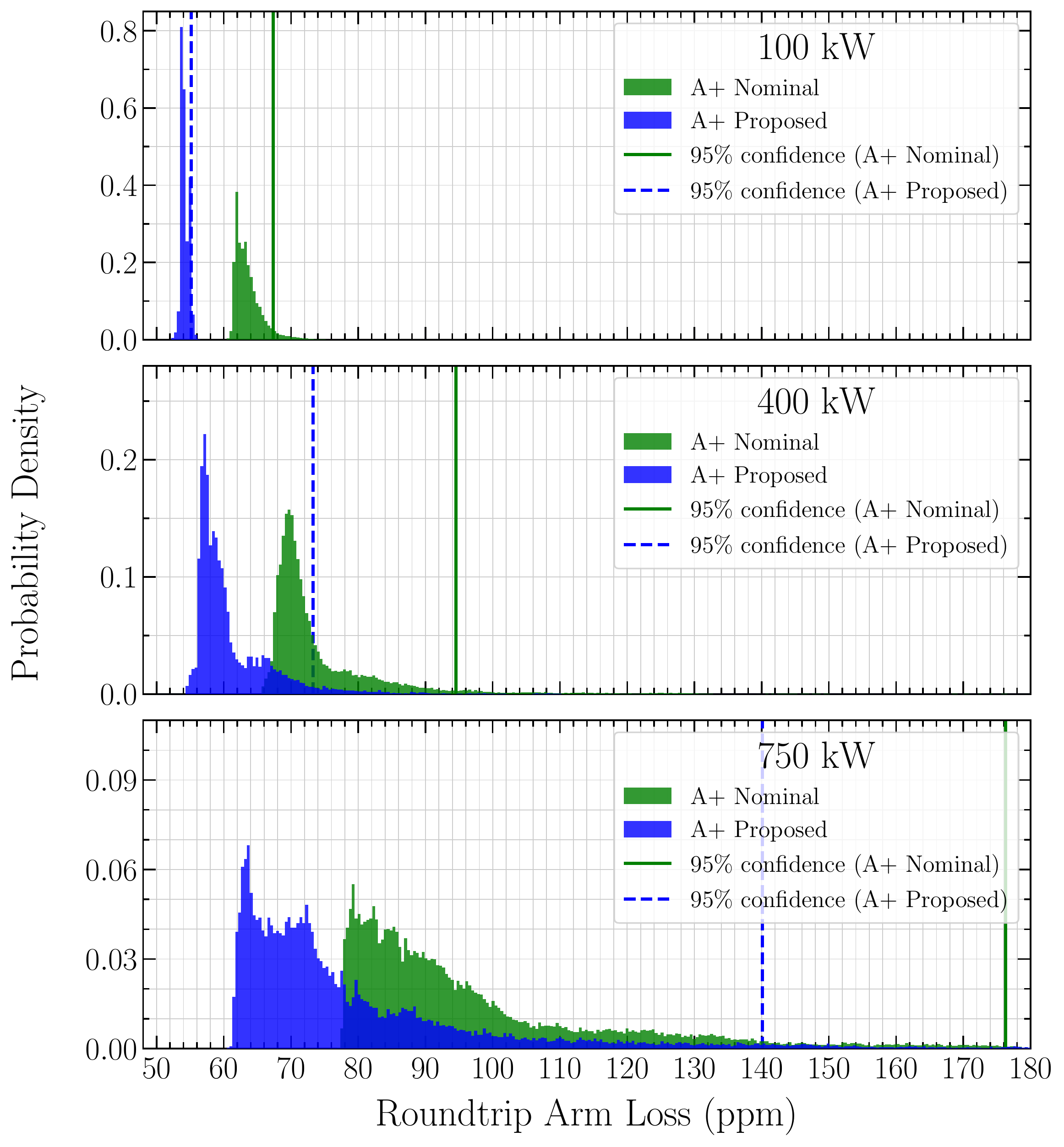}
    \caption{Arm cavity loss distributions due to point absorbers, shown at three different power levels. Each loss distribution represents 1000 trials with a point absorber randomly positioned on each test mass. Uniform coating absorption of 0.3~ppm per test mass, along with optimal ring heater compensation, is assumed.}
    \label{fig:point_absorbers}
\end{figure}

Next, we add random point absorbers to the Monte Carlo simulation and reevaluate the loss performance of both designs. One point absorber is applied to each test mass, randomly positioned in the central 150~mm diameter. The radial and angular coordinates are drawn from uniform distributions, with the radial distribution truncated at 75~mm and the angular distribution spanning the full $360^{\circ}$. Point absorber phase maps are generated using the analytic formalism for thermoelastic surface deformation from Brooks et al.~\cite{Brooks:2021}. We assume a fixed absorptivity chosen so that, at a cavity power of 250~kW, a perfectly centered point absorber absorbs 20~mW of incident power. Fig.~\ref{fig:point_absorbers} shows the roundtrip loss distributions for the fundamental cavity mode under each design, at three different arm power levels. We find the proposed profiles statistically outperform the nominal profiles in all cases.

\section{Signal recycling cavity design}
\label{sec:src}
The single largest source of loss in the Advanced LIGO interferometers is spatial mode-mismatch between optical cavities. Mode-mismatch arises from unintended deviations of the as-built optical system from design. The two folding mirrors of the signal recycling cavity (SRC) are known to be especially sensitive to fabrication and installation errors. Even small perturbations in the curvatures and positions of these optics can result in a significant mode-mismatch with the arm cavities.

The impact of mode-mismatch internal to the interferometer on the observed squeezing is difficult to model analytically. In LIGO, the fundamental optical mode is squeezed in the modal basis defined by a parametric amplifier cavity, which serves as the squeezing source. The cavities of the interferometer each define their own basis of optical modes. As the squeezed state propagates through the interferometer, it is transformed from the modal basis of the squeezing source into the basis of each respective cavity. If the spatial modes of the cavities are imperfectly matched, these basis transformations must mix the optical modes. Since only the fundamental mode in the source basis is squeezed, the higher-order modes carry standard vacuum. Thus, basis mixing from mode-mismatch leads to loss. However, unlike dissipative losses, each modal mixing is coherent and unitary---leading to complex interference effects which can potentially increase squeezing losses. To date, the most detailed analytical treatment of the coherent interactions of transverse modal mixing on squeezed states is given by McCuller {\em et al.}~\cite{McCuller:2021}.

In the present work, we use a numerical simulation to model the squeezing degradation from internal mode-mismatches. The aim of this analysis is to identify an SRC design in LIGO~A+ that is maximally robust to common errors which induce mode-mismatch. We identify the maximally error-tolerant design through a numerical optimization procedure described in \S\ref{sec:optimization}, which employs an evolutionary search algorithm in a parameter space spanning all possible SRC designs. The results of this optimization and a quantitative analysis of the design performance are described in \S\ref{sec:optimization_results}.

\subsection{Optimization procedure}
\label{sec:optimization}
Our objective is to find the optimal values of the radii of curvature and positions of the SRC mirrors, such that deviations from these nominal values minimally degrade the squeezing level observed at the interferometer output. To perform this search, we employ a global-best particle swarm optimization (PSO) algorithm provided by the Pyswarms optimization toolkit~\cite{Miranda:2018}. PSO is an evolutionary search algorithm designed to efficiently explore high-dimensional parameter spaces. Initially, many ``particles,'' each, in our case, representing a candidate optical design, are scattered around the parameter space. Each particle has an associated velocity which determines its position in the parameter space at the following iteration. Its velocity is determined by its best known local position as well as the best positions discovered by other particles. In this way, the entire swarm is iteratively guided toward the global optimum.

The relative ``goodness'' of positions within the parameter space is quantified by a cost function whose value the algorithm seeks to minimize. Primarily, our cost function is designed to penalize SRC designs in which the observed squeezing level is strongly sensitive to perturbations of the SRC parameters. To evaluate the cost function at each particle's position, at each iteration, a \textsc{Finesse} optical simulation~\cite{FinesseRef} is used to compute the partial derivatives of the observed squeezing level with respect to small detunings of each candidate SRC parameter. We detail the \textsc{Finesse} simulation in \S\ref{sec:finesse_model}. Then, in \S\ref{sec:parameters} we describe the set of optical parameters which we optimize, as well as relevant parameter constraints. Finally, in \S\ref{sec:cost_function} we detail the construction of the cost function used to define the optimization objective.

\subsubsection{Optical simulation}
\label{sec:finesse_model}

\begin{figure}[t]
    \centering
    \includegraphics[width=0.48\textwidth]{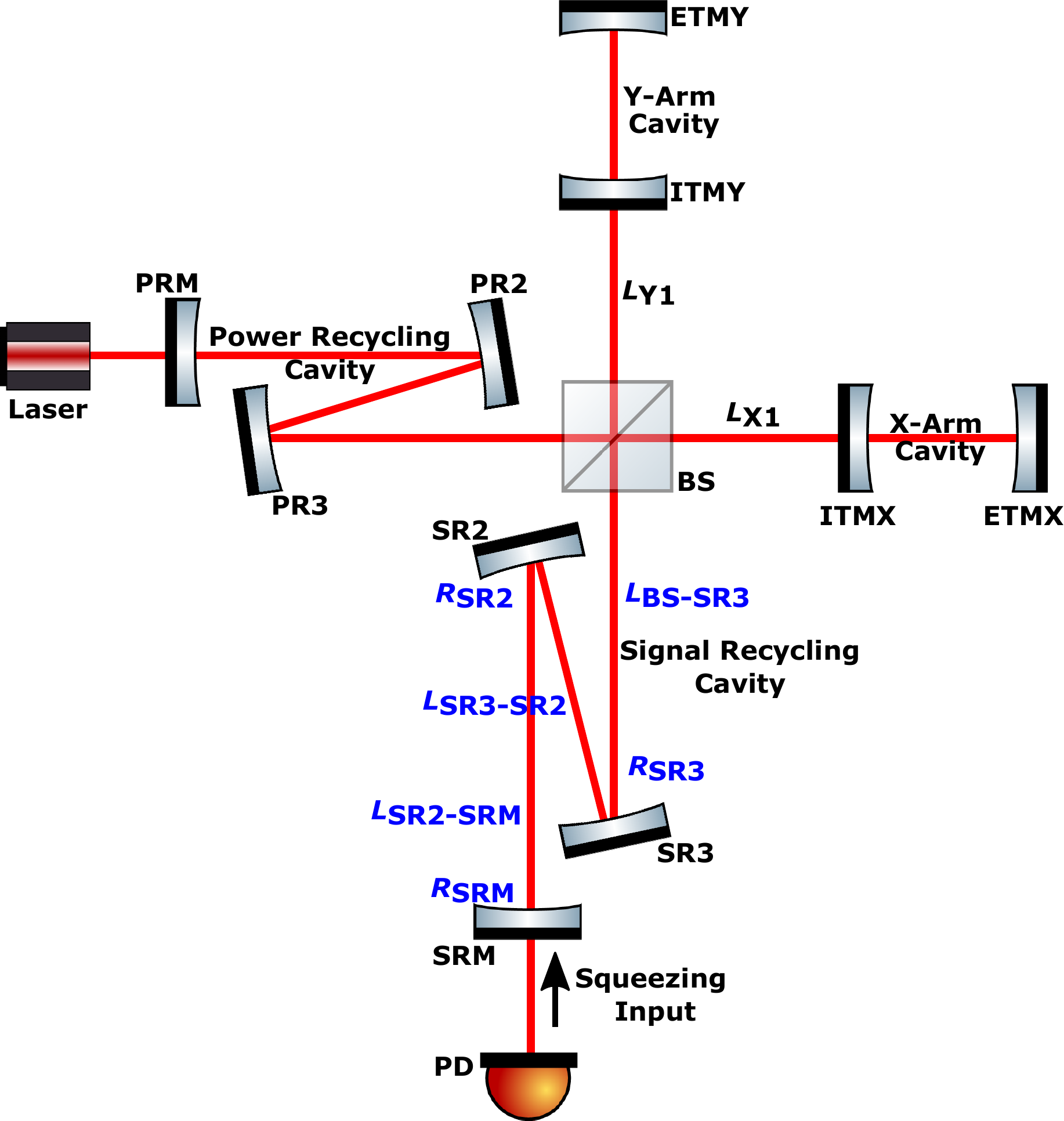}
    \caption{Optical configuration used for simulating a LIGO-like interferometer. All of the distances and radii of curvature are fixed to the nominal LIGO A+ design values, except for the signal recycling cavity parameters which are indicated in blue.}
    \label{fig:setup}
\end{figure}

In our optimization routine, the core compute engine is a \textsc{Finesse} simulation~\cite{FinesseRef} used to analyze the performance of a given SRC design. \textsc{Finesse} is a modal-based optical simulation package widely used for modeling laser cavities, whose modern user interface is provided by the \textsc{Pykat}~\cite{Brown:2020} package. To illustrate the optimization procedure, we adopt a toy interferometer model based on the LIGO A+ design. Its optical layout is shown in Fig.~\ref{fig:setup}. However, for the purpose of this illustration, several simplifying departures from the A+ design are made to reduce the computational cost and complexity:
\begin{itemize}
    \item {\em Frequency-independent squeezing.} Although A+ will use frequency-dependent squeezing~\cite{Fritschel:2020b}, for simplicity we assume a frequency-independent squeezing angle. In principle, our routine can be extended to the frequency-dependent case by jointly optimizing the error tolerance at multiple frequencies.
    \item {\em DC readout.} For signal detection, a bright carrier field must present at the interferometer output port. In Advanced LIGO, this is generated by offsetting the differential arm length $\sim 1$~pm from a dark fringe. In A+, this technique, known as ``DC readout,'' will be replaced by balanced homodye readout~\cite{Fritschel:2014}. As balanced homodyne readout adds considerable complexity, we use DC readout in this simulation.
    \item {\em Output mode-matching.} Inclusion of an output mode cleaner (OMC) adds significant computational cost because, as SRC parameters are detuned, at least two adaptive optics between the SRC and OMC must be continually re-optimized to maintain the mode-matching of the OMC to the arm cavities. Although such a re-tuning of the output mode-matching been previously demonstrated~\cite{Perreca:2020}, for the present simulation we omit the OMC and instead assume a fixed output loss ranging from $5-20\%$.
\end{itemize}

The \textsc{Finesse} simulation starts from a ``nominal'' model (using a provided set of SRC parameters), then individually detunes each SRC parameter from its design value and computes the change in observed squeezing. The parameter detunings introduce a spatial mode-mismatch between the SRC and the arm cavities. Higher-order modes up to order~4 are tracked, which is sufficient given the small size of the parameter detunings. Coincidentally, the mode-mismatch shifts interferometer length degrees of freedom away from their nominal operating points, as well as rotates the squeezing quadrature away from the interferometer readout quadrature. In a real detector, these offsets are zeroed by a combination of control servos and manual optimizations. Thus, it is necessary to implement servos within the \textsc{Finesse} simulation to zero all such ``artifical'' detunings.

To prevent length detunings, we incorporate DC servos for all five length degrees of freedom: the common arm length, differential arm length, power recycling cavity length, Michelson length, and signal recycling cavity length~\cite[see, e.g.,][]{Martynov:2016}. Linear error signals are constructed by injecting 9~MHz and 45~MHz phase modulation sidebands at the interferometer input and measuring the demodulated fields at the symmetric port, antisymmetric port, and a pick-off port inside the power recycling cavity. Every time an SRC parameter is varied, we re-orthogonalize the sensing matrix. Then, to verify the new servo points, we individually detune each length degree of freedom and verify its error signal to be at a zero crossing.

To calculate the observed squeezing level for the detector, we inject a $-14$~dB squeezed vacuum source at the output of the SRC, as shown in Fig.~\ref{fig:setup}. The injected squeezing level is chosen to match that expected for LIGO A+. We then rotate the squeezing angle so as to minimize the quantum (shot) noise level in the interferometer readout channel at 1~kHz. The signal frequency of 1~kHz is chosen to lie in LIGO's high-frequency, shot-noise-dominated band, where optomechanical interactions with the interferometer optics may be neglected. The injected squeezed field is mode-matched to the interferometer arm cavities, rather than to the low-finesse SRC, equivalently to the procedure in use for the real detectors. Every time an SRC parameter is varied, we adjust the input squeezing mode to recover the mode-matching to the arm cavities, then retune the squeezing angle for maximum shot noise reduction.

\subsubsection{Parameters and constraints}
\label{sec:parameters}
During optimization, we allow the lengths and mirror curvatures defining the SRC to vary, while keeping the arm cavities and the power recycling cavity fixed. There are thus six degrees of freedom, as indicated in blue lettering in Fig.~\ref{fig:setup}: the radii of curvature of the SR3, SR2, and SRM mirrors ($R_{\rm  SR3}$, $R_{\rm SR2}$, and $\rm R_{\rm SRM}$, respectively) and the distances between ITMX/Y and SR3, SR3 and SR2, and SR2 and the SRM ($L_{\rm ITM\mbox{-}SR3}$, $L_{\rm SR3\mbox{-}SR2}$, and $L_{\rm SR2\mbox{-}SRM}$, respectively). As shown in Fig.~\ref{fig:setup}, the distance between the ITMs and SR3 is the sum of the distances between the ITMs and beamsplitter ($L_{\rm X1}$ and $L_{\rm Y1}$) and the beamsplitter and SR3 ($L_{\rm BS\mbox{-}SR3}$). To avoid changing the power recycling cavity mode, we allow only the $L_{\rm BS\mbox{-}SR3}$ component to vary.

It is necessary to impose two constraints on the SRC parameters, as described below. In effect, these constraints reduce the dimensionality of the optimization problem from six to four.

\paragraph{Total length.}
For length sensing and control of the SRC, Advanced LIGO relies on the resonance of a 45~MHz phase modulation sideband in this cavity~\cite{Staley:2014}. In order to avoid requiring a major change in the control system, the 45~MHz sideband must remain resonant in the redesigned cavity. Thus, we require the {\em total} SRC length to remain fixed. This reduces the dimensionality of the optimization problem from six to five, via the constraint
\begin{equation}
    \label{eq:lSR1}
    L_{\rm SRC} = L_{\rm ITM\mbox{-}SR3} + L_{\rm SR3\mbox{-}SR2} + L_{\rm SR2\mbox{-}SRM}
\end{equation}
where $L_{\rm SRC}=56.01$~m is the current SRC length.
\\

\paragraph{Mode-matching.}
In order to read out the interferometer signal field through the SRC, the SRC must be mode-matched to the arm cavities. Thus, at the longitudinal location of the ITM reflective surface, $z=z_{\rm ITM}$, we require that the beam parameter of the SRC, $q_{\rm SRC}$, equal that of the arm cavities, $q_{\rm arm}$:
\begin{equation}
    \label{eq:beamParams}
    q_{\rm SRC}\left(z_{\rm ITM}\right) = q_{\rm arm}\left(z_{\rm ITM}\right)
\end{equation}
This mode-matching constraint implies that for one roundtrip traversal through the SRC, starting from the ITM, the ABCD matrix~\cite{Kogelnik:1966} of the SRC must satisfy
\begin{equation}
    \label{eq:curvSR2}
    q_{\rm arm}\left(z_{\rm ITM}\right) = \frac{A \, q_{\rm arm}\left(z_{\rm ITM}\right) + B}{C \, q_{\rm arm}\left(z_{\rm ITM}\right) + D} \;.
\end{equation}
Implicitly, the matrix elements $A$, $B$, $C$, and $D$ are functions of the six SRC design parameters. We numerically solve Eq.~\ref{eq:curvSR2} for $R_{\rm SR2}$ in terms of the other five parameters, further reducing the dimensionality of the optimization problem from five to four.

\subsubsection{Cost function}
\label{sec:cost_function}
The relative performance of competing optical designs is quantified by a cost function, whose value the optimization procedure seeks to minimize. Unlike classical optimization methods, PSO does not use the gradient of the cost function and, thus, does not require it to be differentiable. This allows a high degree of flexibility in construction of the cost function. Primarily, our cost function is designed to penalize SRC designs in which the observed squeezing level is strongly sensitive to perturbations of the SRC parameters ($C_{\rm SQZ}$). Several additional penalties are included to ensure that the cavity is stable ($C_{\rm stable}$), low-loss (due to clipping on the optics; $C_{\rm loss}$), and modally non-degenerate ($C_{\rm HOM}$). The total cost of an SRC design is defined as
\begin{equation}
    \label{eq:cost}
    \textsc{Cost} = C_{\rm SQZ} + C_{\rm stable} + C_{\rm loss} + C_{\rm HOM} \;.
\end{equation}
Each of the terms in Eq.~\ref{eq:cost} is described in detail below.
\\

\paragraph{Squeezing sensitivity ($C_{\rm SQZ}$).}
To quantify the sensitivity of the SRC design to real-world errors, we detune each SRC parameter individually to estimate the partial derivatives of the observed squeezing. Radii of curvature are detuned by $\Delta R = \pm 0.1\%$ and lengths by $\Delta L = \pm 3$~mm. These detunings are chosen to reflect the best achievable fabrication and hand-placement tolerances for LIGO optics, respectively. The design is assigned a cost of
\begin{equation}
    C_{\rm SQZ} \propto \sum_i \left[ \left|\frac{\Delta S_{i,+}}{\Delta x_i}\right| +  \left|\frac{\Delta S_{i,-}}{\Delta x_i}\right| \right] \; ,
\end{equation}
where $\Delta S_{i,\pm}$ is the change in observed squeezing for a positive or negative detuning $\pm \Delta x_i$ of the parameter $x_i$.
\\

\paragraph{Cavity stability ($C_{\rm stable}$).}
To ensure the SRC is a stable optical resonator, we penalize cavities whose $g$ factor is close to the instability limit of $\pm 1$~\cite{Arai:2013}. Designs with stability factors of $\left|g\right| > 0.9$ are assigned a linearly increasing cost of
\begin{equation}
    C_{\rm stable} \propto \left|g\right| \;.
\end{equation}
Designs with stability factors below this threshold are assigned a fixed cost of $C_{\rm stable} = -1$.
\\

\paragraph{Clipping losses ($C_{\rm loss}$).}
A larger beam size on the SRM will result in higher clipping losses further downstream on the output mode-matching optics, which are smaller in diameter. We thus include a penalty to ensure the beam exiting the SRC does not become significantly larger than its present size. At the longitudinal location of the SRM, $z=z_{\rm SRM}$, designs with a Gaussian beam radius $w\left(z_{\rm SRM}\right) > 3$~mm are assigned a linearly increasing cost of
\begin{equation}
    C_{\rm loss} \propto w\left(z_{\rm SRM}\right)  \;.
\end{equation}
Designs with beam sizes below this threshold are assigned a fixed cost of $C_{\rm loss} = -1$.
\\

\paragraph{Modal degeneracy ($C_{\rm HOM}$).}
The presence of higher-order mode (HOM) co-resonances in the SRC can lead to a resonant amplification of scattering losses through a process known as ``mode harming''~\cite{Bochner:2003}. To ensure the SRC is modally non-degenerate, we penalize HOM co-resonances up to order~10. If the roundtrip Guoy phase of the cavity, $\phi_G$, is within $\pm 10\%$ of $2\pi\,/\,n$ for any $n \in \{1,2,..10\}$, the design is assigned a linearly increasing cost of
\begin{equation}
    C_{\rm HOM} \propto 1 -  \frac{\left| \phi_G - 2\pi\,/\,n \right|}{2\pi\,/\,n}  \;.
\end{equation}
Designs with no HOM co-resonances within this threshold are assigned a fixed cost of $C_{\rm HOM} = -1$.

\subsection{Squeezing performance improvement}
\label{sec:optimization_results}

\begin{table}[t]
    \centering
    \def\arraystretch{1.3}
    \begin{tabular*}{\columnwidth}{@{\extracolsep{\stretch{1}}}*{6}{l}@{}}
        \hline\hline
        Parameter & A+ Nominal & A+ Optimal \\ \hline
        SR3 radius of curvature  & 35.97~m & 60.24~m \\
        SR2 radius of curvature & -6.41~m & -4.77~m \\
        SRM radius of curvature & -5.69~m & -56.27~m \\
        Beamsplitter to SR3 length & 19.37~m & 9.97~m \\
        SR3 to SR2 length & 15.44~m & 28.56~m \\
        SR2 to SRM length & 15.76~m & 12.04~m \\
        \hline\hline
    \end{tabular*}
    \caption{Nominal versus optimized signal recycling cavity parameters for LIGO A+.}
    \label{tab:pso_results}
\end{table}

\begin{figure}[t]
    \centering
    \includegraphics[width=0.48\textwidth]{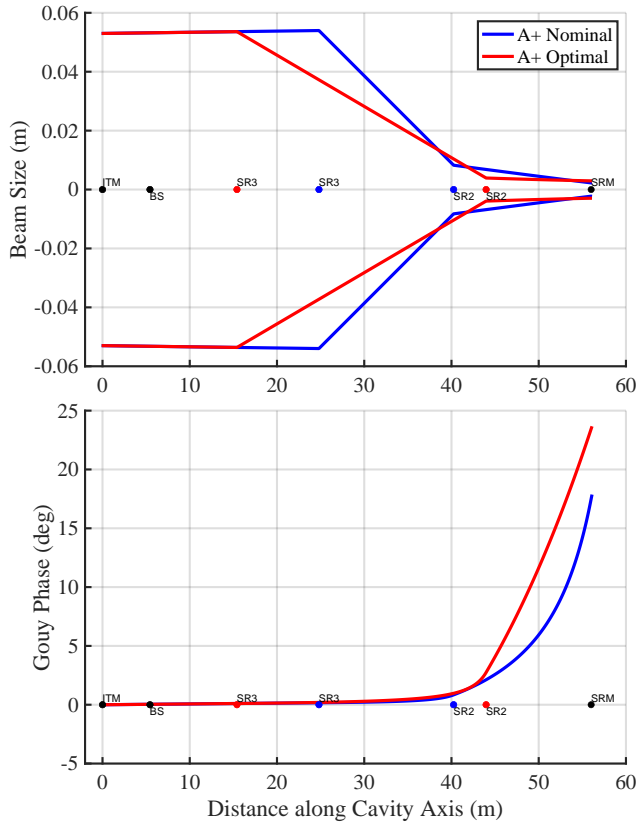}
    \caption{Nominal versus optimized signal recycling cavity designs for LIGO A+. Shown is the Gaussian beam size (top) and the accumulated Gouy phase (bottom) along the cavity axis, from the ITM to the SRM. Of all the optics, only the positions of SR2 and SR3 are allowed to vary.}
    \label{fig:beam_params}
\end{figure}

In this section, we present the most error-tolerant SRC design identified by our optimization routine and characterize its optical performance. Table~\ref{tab:pso_results} lists the optimized SRC parameter values compared to those for the nominal A+ design. The design differences are visualized in Fig.~\ref{fig:beam_params}. The top panel shows the Gaussian beam diameter along the cavity axis, from the ITM to the SRM. The bottom panel shows the accumulated Gouy phase along the same path. As shown, the optimization favors converging the beam more slowly, which is achieved largely by increasing the separation between the SR2 and SR3 telescope mirrors. For the reasons discussed in \S\ref{sec:optimization}, the total SRC length is constrained to remain the same, which fixes the position of the SRM in Fig.~\ref{fig:beam_params}, and a larger beam size at the SRM position is also strongly penalized.

To assess the competitiveness of this candidate design, we analyze its squeezing performance statistically using a Monte Carlo method. With a fixed level of injected squeezing, small random errors are added each of the SRC parameters and the observed squeezing is computed for a large number of trials. The resulting squeezing distributions provide a direct, quantitative comparison of the performance of competing cavity designs. In detail, our procedure is as follows:
\begin{enumerate}
    \item Assume realistic uncertainties in the curvatures and positions of the SRC optics. We assume the uncertainties to be normally distributed with zero mean and a standard deviation of 0.1\% for radius of curvature errors and 3~mm for position errors.
    \item Draw a set of random errors for all six SRC parameters listed in Table~\ref{tab:pso_results}.
    \item Using the \textsc{Finesse} model described in \S\ref{sec:finesse_model}, simulate the interferometer with these perturbed parameters and compute the observed squeezing level. A fixed injected squeezing level of $-14$~dB is assumed.
    \item Repeat the previous steps 2000~times, each time drawing a new set of random parameter errors.
    \item Calculate the probability distribution of observed squeezing across all trials.
\end{enumerate}
Convergence testing with varying numbers of trials has found 2000 to adequately sample the distribution.

\begin{figure}[t]
    \centering
    \includegraphics[width=0.48\textwidth]{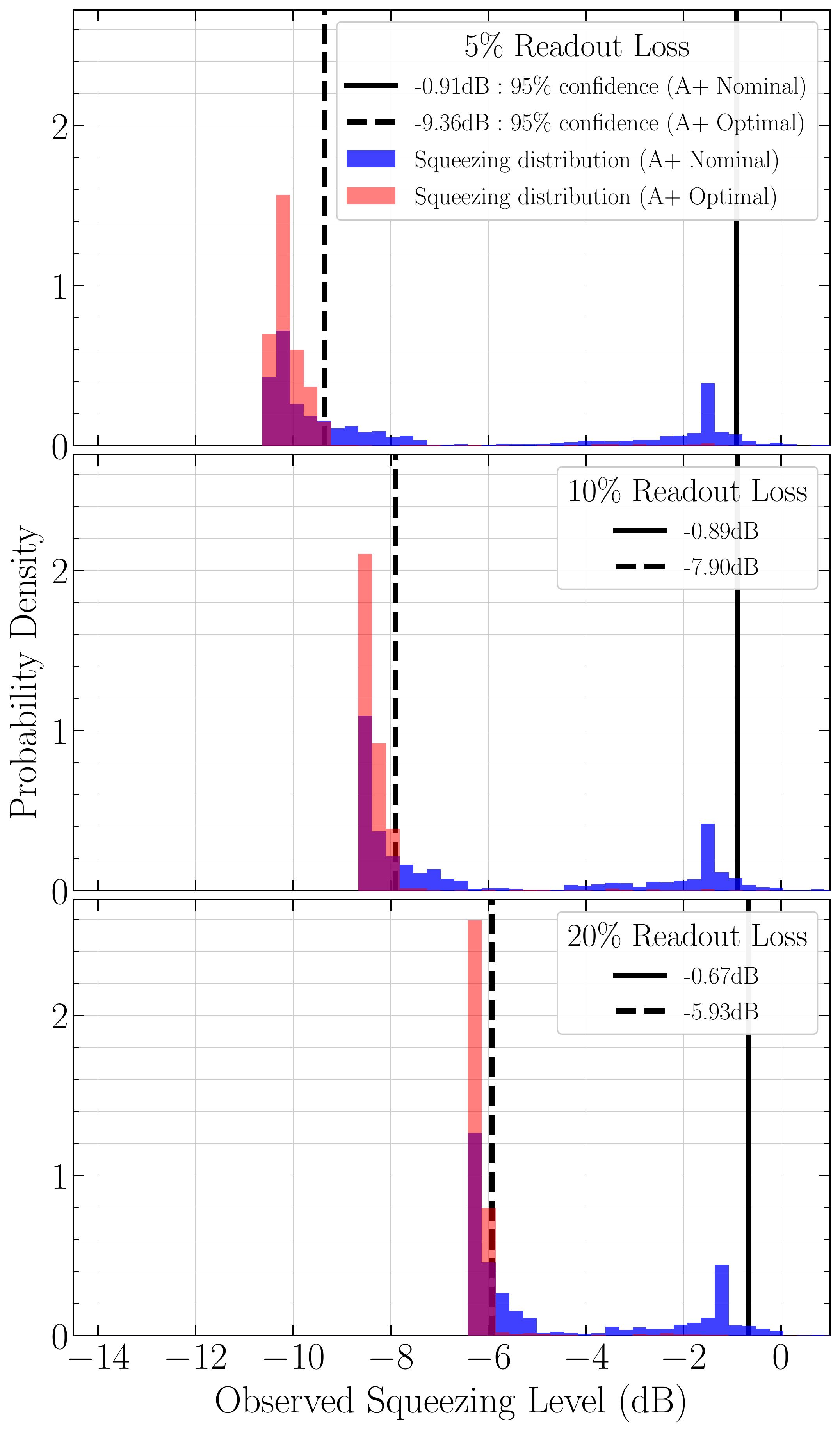}
    \caption{Probability distributions of the squeezing achieved with two different signal recycling cavity (SRC) designs, in the presence of realistic random optical errors. The three panels assume various levels of readout loss ranging from 5\% (top) to 20\% (bottom). In each panel, the black vertical lines indicate the difference in {\em worst} possible outcome, at 95\% confidence, between the two designs.}
    \label{fig:mc_src}
\end{figure}

Fig.~\ref{fig:mc_src} shows the result of this comparative performance analysis for the nominal and optimized SRC designs. The three panels show the probability distributions of observed squeezing under varying levels of readout loss, ranging from 5\% (top panel) to 20\% (bottom panel). This readout loss accounts for attenuation losses due to output mode-mismatch, Faraday isolator insertion loss, optical pick-offs for diagnostic and control purposes, and photodiode quantum inefficiency. At the beginning of LIGO~A+, the readout losses are expected to be similar to the bottommost panel. We find that, in the presence of random optical errors, our optimization procedure results in a significant narrowing of the distribution of possible squeezing outcomes. As shown, this narrowing leads to a modest improvement in the median squeezing level and, at 95\% confidence, a dramatic improvement in the {\em worst} possible outcome (black vertical lines in Fig.~\ref{fig:mc_src}).

The squeezing distributions of the nominal design exhibit a bimodality which arises from two distinct parameter regimes. In each panel, the left peak (corresponding to higher squeezing) overwhelmingly consists of cases in which the SR3 radius of curvature is smaller than intended ($\Delta R_{\rm SR3} < 0.0\%$). On the other hand, the right peak (corresponding to lower squeezing) overwhelmingly consists of large, positive errors in the SR3 radius of curvature ($\Delta R_{\rm SR3} > +0.1\%$). We find no strong correlation in the values of the other five SRC parameters between the two peaks. The strong dependency on $\Delta R_{\rm SR3}$ is reduced but not completely eliminated by the optimization process. In the squeezing distributions of the optimal design, nearly all of the low-squeezing outliers arise from cases of very large, positive error in the SR3 radius of curvature ($\Delta R_{\rm SR3} \ge +0.2\%$).

\begin{figure*}[t]
    \centering
    \includegraphics[width=0.98\textwidth]{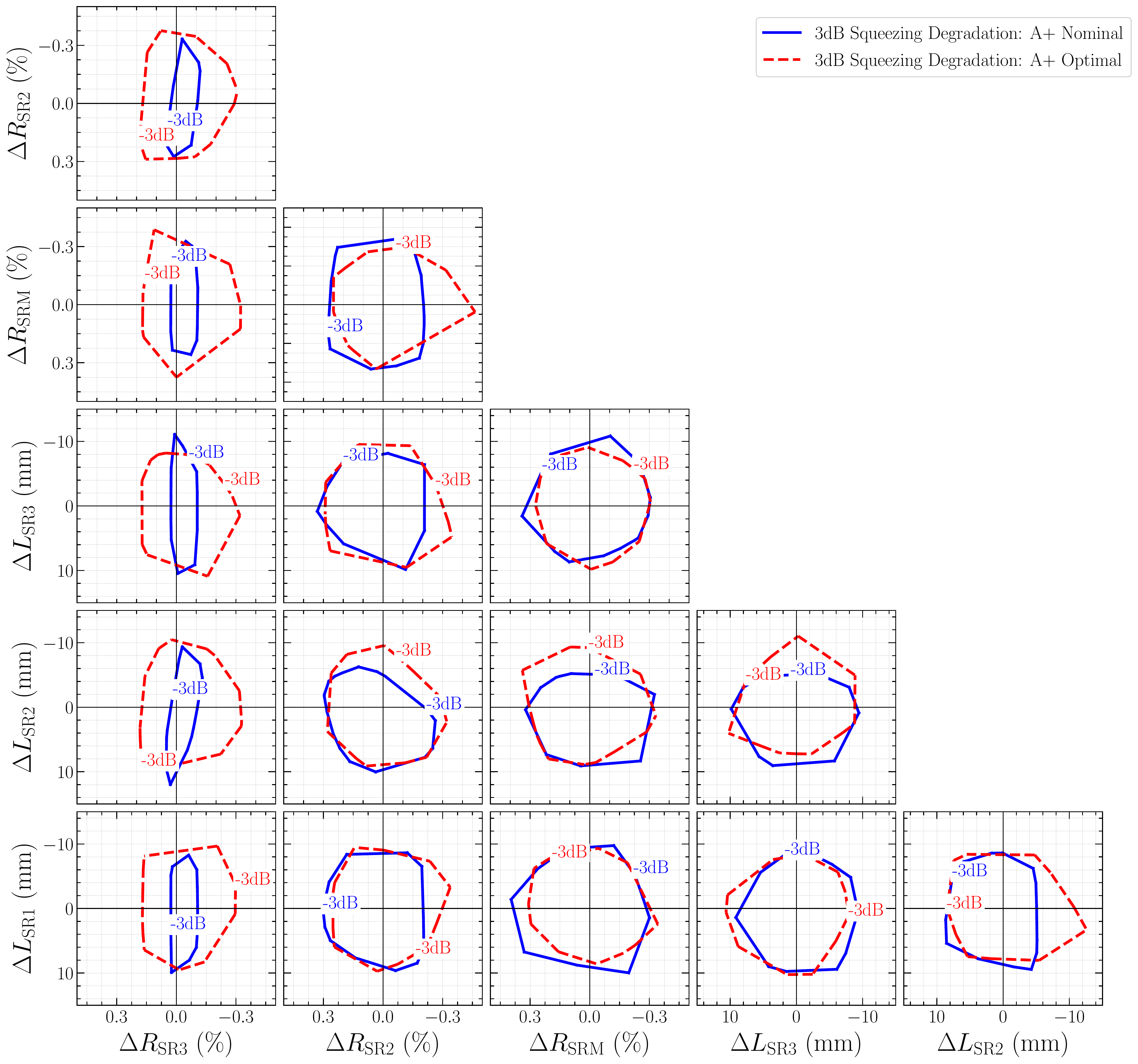}
    \caption{Corner plot comparing the sensitivity to error in pairs of parameters (see Table~\ref{tab:pso_results}) between two signal recycling cavity (SRC) designs. In each panel, the lines represent iso-squeezing contours at which errors in the two parameters degrade the observed squeezing by 3~dB, compared to the unperturbed case (located at the origin). A larger area enclosed by the iso-squeezing contour of one design indicates a greater error tolerance.}
    \label{fig:corner_plt}
\end{figure*}

To understand the extreme sensitivity to errors in the SR3 mirror, and how our optimization process reduces it, we generate a ``corner plot'' by detuning individual pairs of SRC parameters, as shown in Fig.~\ref{fig:corner_plt}. In each panel, the lines represent iso-squeezing contours at which errors in the two parameters degrade the observed squeezing by 3~dB, compared to the unperturbed case (located at the origin). The contours for the nominal SRC design (solid blue lines) and the optimized design (dashed red lines) are overlaid to allow a direct comparison of the parameter sensitivities. A greater error tolerance appears as an increase in the {\em area enclosed} by the iso-squeezing contour (that is, larger parameter errors are required to produce the same degradation in squeezing). As shown, the single largest improvement is a dramatic reduction of sensitivity to errors in $R_{\rm SR3}$.

\section{Conclusions}
\label{sec:conclusions}
The aim of this paper has been to demonstrate the promise of two novel, complementary techniques in optical experiment design:
\begin{enumerate}
    \item {\em Nonspherical mirror surfaces} as solutions to otherwise overconstrained cavity design problems.
    \item {\em Statistics-guided cavity design} for optimal robustness to real-world optical errors.
\end{enumerate}
As a proof of concept, we have performed a two-part optimization of the LIGO A+ design, first modifying the arm cavities for reduced point-absorber-induced loss (see \S\ref{sec:arm}) and then optimizing the SRC for maximum squeezing performance (see \S\ref{sec:src}). Our findings strongly suggest that these techniques can be leveraged to achieve greater performance in current and future gravitational-wave detectors. Both act to minimize internal optical losses, as is critical to achieving megawatt-scale power and high levels of squeezing in third-generation detectors.

Several caveats apply to the results presented herein, which will be the target of future studies. In \S\ref{sec:arm}, the optimal test mass profiles depend critically on the thermal state of the optics. Central heating due to (uniform) coating absorption and thermal compensation applied by the ring heaters induce thermoelastic surface deformations of the same magnitude as the static polish, even at the outer radii. For the purpose of illustration, we assume coating absorption at the same level as for the Advanced LIGO optics. However, the absorptivity of the new coating materials targeted for LIGO A+, $\rm Ti O_2$-doped $\rm Ge O_2$~\cite{Vajente:2021}, is not currently known. Given this uncertainty, an equivalent {\it active} solution is possible in which an annular heating pattern is projected onto the front surface of the test mass near the edge, providing a tunable means of generating surfaces profiles similar to those shown in Fig.~\ref{fig:profiles}.

In \S\ref{sec:src}, an analytical or semi-analytical squeezing model is desirable to cross-validate the numerical results presented in Fig.~\ref{fig:mc_src}. In future work, we aim to develop a semi-analytic mode-scattering model that will enable squeezing calculations to be performed using the FFT-based optical simulation \textsc{SIS}~\cite{Yamamoto:2008}. This will provide an important cross-check of the \textsc{Finesse}-based models. In future work, we also aim to extend the complexity of the optimization in several key ways:
\begin{enumerate}
    \item Include an OMC, to remove sideband power and non-artificially capture output mode-matching losses.
    \item Jointly optimize the quantum noise performance at multiple frequencies, to extend our procedure to frequency-dependent squeezing.
    \item Incorporate additional constraints to generate optical designs compatible with the footprint of the existing LIGO vacuum chambers.
    \item Jointly optimize the power and signal recycling cavities, enabling potential searches for ``mode-healing'' designs~\cite{Bochner:2003} in the presence of point absorbers.
\end{enumerate}
Each of these extensions incurs a considerably higher computational cost than the present work, due either to an increase in the dimensionality of the optimization itself, or by requiring additional sub-optimizations (or servoing) to be performed within each iteration of the optimization. Improved algorithmic efficiency, together with highly parallelized simulations, will be essential to achieving them at the scale of a full gravitational-wave interferometer.

\section*{Acknowledgments}
We are grateful to GariLynn Billingsley for providing test mass mirror maps, Hiro Yamamoto for guidance on using SIS, and Lee McCuller for helpful comments on the squeezing modeling. We would also like to thank the LIGO Laboratory for providing the resources with which to conduct this research, as well as the LIGO SURF program, the National Science Foundation, and the California Institute of Technology for sponsoring the project in part. LIGO was constructed by the California Institute of Technology and Massachusetts Institute of Technology with funding from the National Science Foundation, and operates under Cooperative Agreement No. PHY-1764464. Advanced LIGO was built under Grant No. PHY-0823459. This paper has LIGO Document Number LIGO-P2100184.

\appendix
\section{Arm Cavity Loss by Mode}
\label{sec:loss_by_mode}

\begin{table}[t]
    \centering
    \def\arraystretch{1.5}
    \begin{tabular*}{\columnwidth}{@{\extracolsep{\stretch{1}}}*{6}{l}@{}}
        \hline\hline
        \multicolumn{3}{c}{Mode ${\rm LG}_{p,l}$} & \multicolumn{3}{c}{Roundtrip arm loss} \\
        Order & $p$ & $|l|$ & A+ Nominal & A+ Proposed & Units \\
        \hline
        0 & 0 &  0 & $  76^{+0.7}_{-1.8}$ & $  62^{+0.5}_{-0.3}$& ppm\\
        1 & 0 &  1 & $ 159^{+  4}_{- 15}$ & $ 122^{+  6}_{-  7}$& ppm\\
        2 & 0 &  2 & $ 357^{+ 17}_{- 66}$ & $ 649^{+ 48}_{- 88}$& ppm\\
        2 & 1 &  0 & $ 597^{+ 32}_{-110}$ & $1092^{+ 81}_{-148}$& ppm\\
        3 & 0 &  3 & $ 793^{+ 47}_{-174}$ & $3165^{+220}_{-420}$& ppm\\
        3 & 1 &  1 & $ 1.7^{+0.1}_{-0.4}$ & $ 7.2^{+0.4}_{-0.9}$& ppt\\
        4 & 0 &  4 & $ 2.7^{+0.1}_{-0.4}$ & $  15^{+0.6}_{-1.8}$& ppt\\
        4 & 1 &  2 & $ 7.6^{+0.3}_{-1.1}$ & $  35^{+  1}_{-  4}$& ppt\\
        4 & 2 &  0 & $  10^{+0.4}_{-1.5}$ & $  44^{+  1}_{-  5}$& ppt\\
        5 & 0 &  5 & $  10^{+0.3}_{-1.2}$ & $  42^{+  1}_{-  5}$& ppt\\
        5 & 1 &  3 & $  32^{+  1}_{-  4}$ & $  92^{+  2}_{-  8}$& ppt\\
        5 & 2 &  1 & $  49^{+  2}_{-  6}$ & $ 120^{+  3}_{- 10}$& ppt\\
        6 & 0 &  6 & $  31^{+0.6}_{-2.8}$ & $  80^{+  2}_{-  7}$& ppt\\
        6 & 1 &  4 & $  94^{+  2}_{-  7}$ & $ 160^{+  3}_{- 11}$& ppt\\
        6 & 2 &  2 & $ 150^{+  3}_{- 10}$ & $ 209^{+  3}_{- 11}$& ppt\\
        6 & 3 &  0 & $ 171^{+  3}_{- 11}$ & $ 225^{+  3}_{- 11}$& ppt\\
        7 & 0 &  7 & $  76^{+  2}_{-  6}$ & $ 134^{+  2}_{-  9}$& ppt\\
        7 & 1 &  5 & $ 210^{+  4}_{- 13}$ & $ 261^{+  3}_{-  9}$& ppt\\
        7 & 2 &  3 & $ 312^{+  5}_{- 13}$ & $ 342^{+  3}_{-  8}$& ppt\\
        7 & 3 &  1 & $ 360^{+  6}_{- 13}$ & $ 379^{+  3}_{-  9}$& ppt\\
        8 & 0 &  8 & $ 154^{+  3}_{- 10}$ & $ 224^{+  2}_{-  7}$& ppt\\
        8 & 1 &  6 & $ 348^{+  5}_{- 11}$ & $ 388^{+  3}_{-  8}$& ppt\\
        8 & 2 &  4 & $ 463^{+  6}_{- 12}$ & $ 472^{+  3}_{- 11}$& ppt\\
        8 & 3 &  2 & $ 512^{+  8}_{- 14}$ & $ 511^{+  4}_{- 12}$& ppt\\
        8 & 4 &  0 & $ 526^{+  8}_{- 15}$ & $ 522^{+  4}_{- 12}$& ppt\\

        \hline\hline
    \end{tabular*}
    \caption{Roundtrip arm cavity loss for each Laguerre-Gauss mode up to order~8, shown for both the nominal and proposed test mass profiles. The lower and upper error bars represent the 16th and 84th percentiles of the loss distributions, respectively (see \S\ref{sec:arm_loss_improvement}).}
    \label{tab:losses}
\end{table}

Although the test mass profiles presented in \S\ref{sec:nonspherical_profiles} are designed to shift the resonance frequencies of 7th-order modes (to reduce point absorber scattering), they also achieve a significantly greater dissipation of mode orders 2-7 as shown in Fig.~\ref{fig:loss_v_mode_order}. Enhancing the dissipation of certain modes may be relevant for improving the damping of parametric instabilities in gravitational-wave detectors~\cite{Evans:2015}. Thus, in Table~\ref{tab:losses} we include a breakdown of the roundtrip dissipation per optical mode.

\bibliographystyle{unsrt}
\bibliography{GWreferences}

\begin{thebibliography}{10}

\bibitem{Skenderis:2008}
Kostas Skenderis and Marika Taylor.
\newblock {The fuzzball proposal for black holes}.
\newblock {\em Physics Reports}, 467(4):117 -- 171, 2008.

\bibitem{Cardoso:2017}
Vitor {Cardoso} and Paolo {Pani}.
\newblock {Tests for the existence of black holes through gravitational wave
  echoes}.
\newblock {\em Nature Astronomy}, 1:586--591, September 2017.

\bibitem{Brustein:2018}
Ram Brustein and A.~J.~M. Medved.
\newblock {Quantum hair of black holes out of equilibrium}.
\newblock {\em Phys. Rev. D}, 97:044035, February 2018.

\bibitem{Markakis:2009}
C~Markakis, J~S Read, M~Shibata, K~Ury{\={u}}, J~D~E Creighton, J~L Friedman,
  and B~D Lackey.
\newblock {Neutron star equation of state via gravitational wave observations}.
\newblock {\em Journal of Physics: Conference Series}, 189:012024, October
  2009.

\bibitem{Read:2013}
Jocelyn~S. Read, Luca Baiotti, Jolien D.~E. Creighton, John~L. Friedman, Bruno
  Giacomazzo, Koutarou Kyutoku, Charalampos Markakis, Luciano Rezzolla, Masaru
  Shibata, and Keisuke Taniguchi.
\newblock {Matter effects on binary neutron star waveforms}.
\newblock {\em Phys. Rev. D}, 88:044042, August 2013.

\bibitem{Chen:2018}
Hsin-Yu {Chen}, Maya {Fishbach}, and Daniel~E. {Holz}.
\newblock {{A two per cent Hubble constant measurement from standard sirens
  within five years}}.
\newblock {\em Nature}, 562(7728):545--547, October 2018.

\bibitem{Riess:2019}
Adam~G. Riess, Stefano Casertano, Wenlong Yuan, Lucas~M. Macri, and Dan
  Scolnic.
\newblock {Large Magellanic Cloud} cepheid standards provide a 1{\%} foundation
  for the determination of the {Hubble} constant and stronger evidence for
  physics beyond {$\Lambda$}{CDM}.
\newblock {\em The Astrophysical Journal}, 876(1):85, May 2019.

\bibitem{Buikema:2020}
A.~Buikema, C.~Cahillane, G.~L. Mansell, C.~D. Blair, R.~Abbott, C.~Adams,
  R.~X. Adhikari, A.~Ananyeva, S.~Appert, K.~Arai, J.~S. Areeda, Y.~Asali,
  S.~M. Aston, C.~Austin, A.~M. Baer, M.~Ball, S.~W. Ballmer, S.~Banagiri,
  D.~Barker, L.~Barsotti, J.~Bartlett, B.~K. Berger, J.~Betzwieser,
  D.~Bhattacharjee, G.~Billingsley, S.~Biscans, R.~M. Blair, N.~Bode,
  P.~Booker, R.~Bork, A.~Bramley, A.~F. Brooks, D.~D. Brown, K.~C. Cannon,
  X.~Chen, A.~A. Ciobanu, F.~Clara, S.~J. Cooper, K.~R. Corley, S.~T.
  Countryman, P.~B. Covas, D.~C. Coyne, L.~E.~H. Datrier, D.~Davis,
  C.~Di~Fronzo, K.~L. Dooley, J.~C. Driggers, P.~Dupej, S.~E. Dwyer, A.~Effler,
  T.~Etzel, M.~Evans, T.~M. Evans, J.~Feicht, A.~Fernandez-Galiana,
  P.~Fritschel, V.~V. Frolov, P.~Fulda, M.~Fyffe, J.~A. Giaime, K.~D. Giardina,
  P.~Godwin, E.~Goetz, S.~Gras, C.~Gray, R.~Gray, A.~C. Green, E.~K. Gustafson,
  R.~Gustafson, J.~Hanks, J.~Hanson, T.~Hardwick, R.~K. Hasskew, M.~C. Heintze,
  A.~F. Helmling-Cornell, N.~A. Holland, J.~D. Jones, S.~Kandhasamy, S.~Karki,
  M.~Kasprzack, K.~Kawabe, N.~Kijbunchoo, P.~J. King, J.~S. Kissel, Rahul
  Kumar, M.~Landry, B.~B. Lane, B.~Lantz, M.~Laxen, Y.~K. Lecoeuche,
  J.~Leviton, J.~Liu, M.~Lormand, A.~P. Lundgren, R.~Macas, M.~MacInnis, D.~M.
  Macleod, S.~M\'arka, Z.~M\'arka, D.~V. Martynov, K.~Mason, T.~J. Massinger,
  F.~Matichard, N.~Mavalvala, R.~McCarthy, D.~E. McClelland, S.~McCormick,
  L.~McCuller, J.~McIver, T.~McRae, G.~Mendell, K.~Merfeld, E.~L. Merilh,
  F.~Meylahn, T.~Mistry, R.~Mittleman, G.~Moreno, C.~M. Mow-Lowry, S.~Mozzon,
  A.~Mullavey, T.~J.~N. Nelson, P.~Nguyen, L.~K. Nuttall, J.~Oberling,
  Richard~J. Oram, B.~O'Reilly, C.~Osthelder, D.~J. Ottaway, H.~Overmier, J.~R.
  Palamos, W.~Parker, E.~Payne, A.~Pele, R.~Penhorwood, C.~J. Perez,
  M.~Pirello, H.~Radkins, K.~E. Ramirez, J.~W. Richardson, K.~Riles, N.~A.
  Robertson, J.~G. Rollins, C.~L. Romel, J.~H. Romie, M.~P. Ross, K.~Ryan,
  T.~Sadecki, E.~J. Sanchez, L.~E. Sanchez, T.~R. Saravanan, R.~L. Savage,
  D.~Schaetzl, R.~Schnabel, R.~M.~S. Schofield, E.~Schwartz, D.~Sellers,
  T.~Shaffer, D.~Sigg, B.~J.~J. Slagmolen, J.~R. Smith, S.~Soni, B.~Sorazu,
  A.~P. Spencer, K.~A. Strain, L.~Sun, M.~J. Szczepa\ifmmode~\acute{n}\else
  \'{n}\fi{}czyk, M.~Thomas, P.~Thomas, K.~A. Thorne, K.~Toland, C.~I. Torrie,
  G.~Traylor, M.~Tse, A.~L. Urban, G.~Vajente, G.~Valdes, D.~C. Vander-Hyde,
  P.~J. Veitch, K.~Venkateswara, G.~Venugopalan, A.~D. Viets, T.~Vo,
  C.~Vorvick, M.~Wade, R.~L. Ward, J.~Warner, B.~Weaver, R.~Weiss, C.~Whittle,
  B.~Willke, C.~C. Wipf, L.~Xiao, H.~Yamamoto, Hang Yu, Haocun Yu, L.~Zhang,
  M.~E. Zucker, and J.~Zweizig.
\newblock {Sensitivity and performance of the Advanced LIGO detectors in the
  third observing run}.
\newblock {\em Phys. Rev. D}, 102:062003, September 2020.

\bibitem{Caves:1980}
Carlton Caves, Kip Thorne, Ronald Drever, Vernon Sandberg, and Mark Zimmermann.
\newblock {On the measurement of a weak classical force coupled to a
  quantum-mechanical oscillator. I. Issues of principle}.
\newblock {\em Reviews of Modern Physics}, 52(2):341--392, April 1980.

\bibitem{Caves:1981}
Carlton~M. Caves.
\newblock {Quantum-mechanical noise in an interferometer}.
\newblock {\em Phys. Rev. D}, 23:1693--1708, April 1981.

\bibitem{Brooks:2021}
Aidan~F. Brooks, Gabriele Vajente, Hiro Yamamoto, Rich Abbott, Carl Adams,
  Rana~X. Adhikari, Alena Ananyeva, Stephen Appert, Koji Arai, Joseph~S.
  Areeda, Yasmeen Asali, Stuart~M. Aston, Corey Austin, Anne~M. Baer, Matthew
  Ball, Stefan~W. Ballmer, Sharan Banagiri, David Barker, Lisa Barsotti,
  Jeffrey Bartlett, Beverly~K. Berger, Joseph Betzwieser, Dripta Bhattacharjee,
  Garilynn Billingsley, Sebastien Biscans, Carl~D. Blair, Ryan~M. Blair, Nina
  Bode, Phillip Booker, Rolf Bork, Alyssa Bramley, Daniel~D. Brown, Aaron
  Buikema, Craig Cahillane, Kipp~C. Cannon, Huy~Tuong Cao, Xu~Chen, Alexei~A.
  Ciobanu, Filiberto Clara, Camilla Compton, Sam~J. Cooper, Kenneth~R. Corley,
  Stefan~T. Countryman, Pep~B. Covas, Dennis~C. Coyne, Laurence~E. Datrier,
  Derek Davis, Chiara~D. Difronzo, Katherine~L. Dooley, Jenne~C. Driggers,
  Peter Dupej, Sheila~E. Dwyer, Anamaria Effler, Todd Etzel, Matthew Evans,
  Tom~M. Evans, Jon Feicht, Alvaro Fernandez-Galiana, Peter Fritschel,
  Valery~V. Frolov, Paul Fulda, Michael Fyffe, Joe~A. Giaime, Dwayne~D.
  Giardina, Patrick Godwin, Evan Goetz, Slawomir Gras, Corey Gray, Rachel Gray,
  Anna~C. Green, Anchal Gupta, Eric~K. Gustafson, Dick Gustafson, Evan Hall,
  Jonathan Hanks, Joe Hanson, Terra Hardwick, Raine~K. Hasskew, Matthew~C.
  Heintze, Adrian~F. Helmling-Cornell, Nathan~A. Holland, Kiamu Izmui, Wenxuan
  Jia, Jeff~D. Jones, Shivaraj Kandhasamy, Sudarshan Karki, Marie Kasprzack,
  Keita Kawabe, Nutsinee Kijbunchoo, Peter~J. King, Jeffrey~S. Kissel, Rahul
  Kumar, Michael Landry, Benjamin~B. Lane, Brian Lantz, Michael Laxen,
  Yannick~K. Lecoeuche, Jessica Leviton, Liu Jian, Marc Lormand, Andrew~P.
  Lundgren, Ronaldas Macas, Myron Macinnis, Duncan~M. Macleod, Georgia~L.
  Mansell, Szabolcs Marka, Zsuzsanna Marka, Denis~V. Martynov, Ken Mason,
  Thomas~J. Massinger, Fabrice Matichard, Nergis Mavalvala, Richard McCarthy,
  David~E. McClelland, Scott McCormick, Lee McCuller, Jessica McIver, Terry
  McRae, Gregory Mendell, Kara Merfeld, Edmond~L. Merilh, Fabian Meylahn,
  Timesh Mistry, Richard Mittleman, Gerardo Moreno, Conor~M. Mow-Lowry, Simone
  Mozzon, Adam Mullavey, Timothy~J. Nelson, Philippe Nguyen, Laura~K. Nuttall,
  Jason Oberling, Richard~J. Oram, Charles Osthelder, David~J. Ottaway, Harry
  Overmier, Jordan~R. Palamos, William Parker, Ethan Payne, Arnaud Pele, Reilly
  Penhorwood, Carlos~J. Perez, Marc Pirello, Hugh Radkins, Karla~E. Ramirez,
  Jonathan~W. Richardson, Keith Riles, Norna~A. Robertson, Jameson~G. Rollins,
  Chandra~L. Romel, Janeen~H. Romie, Michael~P. Ross, Kyle Ryan, Travis
  Sadecki, Eduardo~J. Sanchez, Luis~E. Sanchez, Saravanan~R.
  Tiruppatturrajamanikkam, Richard~L. Savage, Dean Schaetzl, Roman Schnabel,
  Robert~M. Schofield, Eyal Schwartz, Danny Sellers, Thomas Shaffer, Daniel
  Sigg, Bram~J. Slagmolen, Joshua~R. Smith, Siddharth Soni, Borja Sorazu,
  Andrew~P. Spencer, Ken~A. Strain, Ling Sun, Marek~J. Szczepanczyk, Michael
  Thomas, Patrick Thomas, Keith~A. Thorne, Karl Toland, Calum~I. Torrie, Gary
  Traylor, Maggie Tse, Alexander~L. Urban, Guillermo Valdes, Daniel~C.
  Vander-Hyde, Peter~J. Veitch, Krishna Venkateswara, Gautam Venugopalan,
  Aaron~D. Viets, Thomas Vo, Cheryl Vorvick, Madeline Wade, Robert~L. Ward, Jim
  Warner, Betsy Weaver, Rainer Weiss, Chris Whittle, Benno Willke,
  Christopher~C. Wipf, Liting Xiao, Hang Yu, Haocun Yu, Liyuan Zhang,
  Michael~E. Zucker, and John Zweizig.
\newblock Point absorbers in {Advanced LIGO}.
\newblock {\em Appl. Opt.}, 60(13):4047--4063, May 2021.

\bibitem{Tse:2019}
M.~Tse, Haocun Yu, N.~Kijbunchoo, A.~Fernandez-Galiana, P.~Dupej, L.~Barsotti,
  C.~D. Blair, D.~D. Brown, S.~E. Dwyer, A.~Effler, M.~Evans, P.~Fritschel,
  V.~V. Frolov, A.~C. Green, G.~L. Mansell, F.~Matichard, N.~Mavalvala, D.~E.
  McClelland, L.~McCuller, T.~McRae, J.~Miller, A.~Mullavey, E.~Oelker, I.~Y.
  Phinney, D.~Sigg, B.~J.~J. Slagmolen, T.~Vo, R.~L. Ward, C.~Whittle,
  R.~Abbott, C.~Adams, R.~X. Adhikari, A.~Ananyeva, S.~Appert, K.~Arai, J.~S.
  Areeda, Y.~Asali, S.~M. Aston, C.~Austin, A.~M. Baer, M.~Ball, S.~W. Ballmer,
  S.~Banagiri, D.~Barker, J.~Bartlett, B.~K. Berger, J.~Betzwieser,
  D.~Bhattacharjee, G.~Billingsley, S.~Biscans, R.~M. Blair, N.~Bode,
  P.~Booker, R.~Bork, A.~Bramley, A.~F. Brooks, A.~Buikema, C.~Cahillane, K.~C.
  Cannon, X.~Chen, A.~A. Ciobanu, F.~Clara, S.~J. Cooper, K.~R. Corley, S.~T.
  Countryman, P.~B. Covas, D.~C. Coyne, L.~E.~H. Datrier, D.~Davis,
  C.~Di~Fronzo, J.~C. Driggers, T.~Etzel, T.~M. Evans, J.~Feicht, P.~Fulda,
  M.~Fyffe, J.~A. Giaime, K.~D. Giardina, P.~Godwin, E.~Goetz, S.~Gras,
  C.~Gray, R.~Gray, Anchal Gupta, E.~K. Gustafson, R.~Gustafson, J.~Hanks,
  J.~Hanson, T.~Hardwick, R.~K. Hasskew, M.~C. Heintze, A.~F. Helmling-Cornell,
  N.~A. Holland, J.~D. Jones, S.~Kandhasamy, S.~Karki, M.~Kasprzack, K.~Kawabe,
  P.~J. King, J.~S. Kissel, Rahul Kumar, M.~Landry, B.~B. Lane, B.~Lantz,
  M.~Laxen, Y.~K. Lecoeuche, J.~Leviton, J.~Liu, M.~Lormand, A.~P. Lundgren,
  R.~Macas, M.~MacInnis, D.~M. Macleod, S.~M\'arka, Z.~M\'arka, D.~V. Martynov,
  K.~Mason, T.~J. Massinger, R.~McCarthy, S.~McCormick, J.~McIver, G.~Mendell,
  K.~Merfeld, E.~L. Merilh, F.~Meylahn, T.~Mistry, R.~Mittleman, G.~Moreno,
  C.~M. Mow-Lowry, S.~Mozzon, T.~J.~N. Nelson, P.~Nguyen, L.~K. Nuttall,
  J.~Oberling, R.~J. Oram, B.~O'Reilly, C.~Osthelder, D.~J. Ottaway,
  H.~Overmier, J.~R. Palamos, W.~Parker, E.~Payne, A.~Pele, C.~J. Perez,
  M.~Pirello, H.~Radkins, K.~E. Ramirez, J.~W. Richardson, K.~Riles, N.~A.
  Robertson, J.~G. Rollins, C.~L. Romel, J.~H. Romie, M.~P. Ross, K.~Ryan,
  T.~Sadecki, E.~J. Sanchez, L.~E. Sanchez, T.~R. Saravanan, R.~L. Savage,
  D.~Schaetzl, R.~Schnabel, R.~M.~S. Schofield, E.~Schwartz, D.~Sellers, T.~J.
  Shaffer, J.~R. Smith, S.~Soni, B.~Sorazu, A.~P. Spencer, K.~A. Strain,
  L.~Sun, M.~J. Szczepa\ifmmode~\acute{n}\else \'{n}\fi{}czyk, M.~Thomas,
  P.~Thomas, K.~A. Thorne, K.~Toland, C.~I. Torrie, G.~Traylor, A.~L. Urban,
  G.~Vajente, G.~Valdes, D.~C. Vander-Hyde, P.~J. Veitch, K.~Venkateswara,
  G.~Venugopalan, A.~D. Viets, C.~Vorvick, M.~Wade, J.~Warner, B.~Weaver,
  R.~Weiss, B.~Willke, C.~C. Wipf, L.~Xiao, H.~Yamamoto, M.~J. Yap, Hang Yu,
  L.~Zhang, M.~E. Zucker, and J.~Zweizig.
\newblock Quantum-enhanced {Advanced LIGO} detectors in the era of
  gravitational-wave astronomy.
\newblock {\em Phys. Rev. Lett.}, 123:231107, December 2019.

\bibitem{Ciriolo:2017}
Anna~G. Ciriolo, Matteo Negro, Michele Devetta, Eugenio Cinquanta, Davide
  Faccialà, Aditya Pusala, Sandro De~Silvestri, Salvatore Stagira, and
  Caterina Vozzi.
\newblock Optical parametric amplification techniques for the generation of
  high-energy few-optical-cycles {IR} pulses for strong field applications.
\newblock {\em Applied Sciences}, 7(3), 2017.

\bibitem{Miao:2019}
Haixing Miao, Nicolas~D. Smith, and Matthew Evans.
\newblock Quantum limit for laser interferometric gravitational-wave detectors
  from optical dissipation.
\newblock {\em Phys. Rev. X}, 9:011053, March 2019.

\bibitem{Perreca:2020}
Antonio Perreca, Aidan~F. Brooks, Jonathan~W. Richardson, Daniel T\"oyr\"a, and
  Rory Smith.
\newblock {Analysis and visualization of the output mode-matching requirements
  for squeezing in Advanced LIGO and future gravitational wave detectors}.
\newblock {\em Phys. Rev. D}, 101:102005, May 2020.

\bibitem{Reitze:2019}
David Reitze, Rana~X Adhikari, Stefan Ballmer, Barry Barish, Lisa Barsotti,
  GariLynn Billingsley, Duncan~A. Brown, Yanbei Chen, Dennis Coyne, Robert
  Eisenstein, Matthew Evans, Peter Fritschel, Evan~D. Hall, Albert Lazzarini,
  Geoffrey Lovelace, Jocelyn Read, B.~S. Sathyaprakash, David Shoemaker, Joshua
  Smith, Calum Torrie, Salvatore Vitale, Rainer Weiss, Christopher Wipf, and
  Michael Zucker.
\newblock {Cosmic Explorer}: The {U.S}. contribution to gravitational-wave
  astronomy beyond {LIGO}.
\newblock {\em Bulletin of the AAS}, 51(7), September 2019.

\bibitem{Abernathy:2011}
M~Abernathy, F~Acernese, P~Ajith, B~Allen, P~Amaro-Seoane, N~Andersson,
  S~Aoudia, P~Astone, B~Krishnan, L~Barack, F~Barone, B~Barr, M~Barsuglia,
  M~Bassan, R~Bassiri, M~Beker, N~Beveridge, M~Bizouard, C~Bond, S~Bose,
  L~Bosi, S~Braccini, C~Bradaschia, M~Britzger, F~Brueckner, T~Bulik,
  HJ~Bulten, O~Burmeister, E~Calloni, P~Campsie, L~Carbone, G~Cella,
  E~Chalkley, E~Chassande-Mottin, S~Chelkowski, A~Chincarini, A~DiCintio,
  J~Clark, E~Coccia, CN~Colacino, J~Colas, A~Colla, A~Corsi, A~Cumming,
  L~Cunningham, E~Cuoco, S~Danilishin, K~Danzmann, E~Daw, R~De Salvo,
  W~DelPozzo, T~Dent, R~DeRosa, L~Di Fiore, M~Di~Paolo Emilio, A~Di Virgilio,
  A~Dietz, M~Doets, J~Dueck, M~Edwards, V~Fafone, S~Fairhurst, P~Falferi,
  M~Favata, V~Ferrari, F~Ferrini, F~Fidecaro, R~Flaminio, J~Franc, F~Frasconi,
  A~Freise, D~Friedrich, P~Fulda, J~Gair, M~Galimberti, G~Gemme, E~Genin,
  A~Gennai, A~Giazotto, K~Glampedakis, R~Gouaty, C~Graef, W~Graham, M~Granata,
  H~Grote, G~Guidi, J~Hallam, G~Hammond, M~Hannam, J~Harms, K~Haughian,
  I~Hawke, D~Heinert, M~Hendry, I~Heng, E~Hennes, S~Hild, J~Hough, D~Huet,
  S~Husa, S~Huttner, B~Iyer, I~Jones, G~Jones, I~Kamaretsos, C~Kant Mishra,
  F~Kawazoe, F~Khalili, B~Kley, K~Kokeyama, K~Kokkotas, S~Kroker, R~Kumar,
  K~Kuroda, B~Lagrange, N~Lastzka, TGF Li, M~Lorenzini, G~Losurdo, H~Luck,
  E~Majorana, V~Malvezzi, I~Mandel, V~Mandic, S~Marka, F~Marin, F~Marion,
  J~Marque, I~Martin, D~McLeod, D~Mckechan, M~Mehmet, C~Michel, Y~Minenkov,
  N~Morgado, A~Morgia, S~Mosca, L~Moscatelli, B~Mours, H~Muller-Ebhardt,
  P~Murray, L~Naticchioni, R~Nawrodt, J~Nelson, R~O'Shaughnessy, CD~Ott,
  C~Palomba, A~Paoli, G~Parguez, A~Pasqualetti, R~Passaquieti, D~Passuello,
  M~Perciballi, F~Piergiovanni, L~Pinard, M~Pitkin, W~Plastino, M~Plissi,
  R~Poggiani, P~Popolizio, E~Porter, M~Prato, G~Prodi, M~Punturo, P~Puppo,
  D~Rabeling, I~Racz, P~Rapagnani, V~Re, J~Read, T~Regimbau, H~Rehbein, S~Reid,
  L~Rezzolla, F~Ricci, F~Richard, A~Rocchi, R~Romano, S~Rowan, A~Rudiger,
  A~Samblowski, L~Santamaria, B~Sassolas, B~Sathyaprakash, R~Schilling,
  P~Schmidt, R~Schnabel, B~Schutz, C~Schwarz, J~Scott, P~Seidel, AM~Sintes,
  K~Somiya, CF~Sopuerta, B~Sorazu, F~Speirits, L~Storchi, K~Strain, S~Strigin,
  P~Sutton, S~Tarabrin, B~Taylor, A~Thurin, K~Tokmakov, M~Tonelli, H~Tourneer,
  R~Vaccarone, H~Vahlbruch, JFJ van~den Brand, C~Van Den~Broeck, S~van~der
  Putten, M~van Veggel, A~Vecchio, J~Veitch, F~Vetrano, A~Vicere, S~Vyatchanin,
  P~Webels, B~Willke, W~Winkler, G~Woan, K~Wojcik, A~Woodcraft, and K~Yamamoto.
\newblock {Einstein Gravitational Wave Telescope} conceptual design study.
\newblock ET Technical Report ET-0106C-10, June 2011.

\bibitem{Fritschel:2020b}
Peter Fritschel, Stuart Reid, Gabriele Vajente, Giles Hammond, Daniel Brown,
  Haixing Miao, and Volker Quetschke.
\newblock Instrument science white paper 2020.
\newblock LIGO Technical Report LIGO-T2000407-v3, 2020.

\bibitem{Vajente:2014}
G.~Vajente.
\newblock {In situ correction of mirror surface to reduce round-trip losses in
  Fabry-Perot cavities}.
\newblock {\em Appl. Opt.}, 53(7):1459--1465, March 2014.

\bibitem{Vajente:2021}
Gabriele Vajente, Le~Yang, Aaron Davenport, Mariana Fazio, Alena Ananyeva,
  Liyuan Zhang, Garilynn Billingsley, Kiran Prasai, Ashot Markosyan, Riccardo
  Bassiri, Martin~M. Fejer, Martin Chicoine, Francois Schiettekatte, and
  Carmen~S. Menoni.
\newblock {Low thermal noise TiO2-doped GeO2 coatings for high sensitivity
  gravitational wave interferometers}.
\newblock LIGO Technical Report LIGO-P2100075, 2021.
\newblock {\it (in preparation)}.

\bibitem{Amato:2019}
Alex Amato, Silvana Terreni, Vincent Dolique, Dani{\`{e}}le Forest, Gianluca
  Gemme, Massimo Granata, Lorenzo Mereni, Christophe Michel, Laurent Pinard,
  Benoit Sassolas, Julien Teillon, Gianpietro Cagnoli, and Maurizio Canepa.
\newblock Optical properties of high-quality oxide coating materials used in
  gravitational-wave advanced detectors.
\newblock {\em Journal of Physics: Materials}, 2(3):035004, jun 2019.

\bibitem{Granata:2020}
M~Granata, A~Amato, L~Balzarini, M~Canepa, J~Degallaix, D~Forest, V~Dolique,
  L~Mereni, C~Michel, L~Pinard, B~Sassolas, J~Teillon, and G~Cagnoli.
\newblock Amorphous optical coatings of present gravitational-wave
  interferometers.
\newblock {\em Classical and Quantum Gravity}, 37(9):095004, apr 2020.

\bibitem{Brooks:2016}
Aidan~F. Brooks, Benjamin Abbott, Muzammil~A. Arain, Giacomo Ciani, Ayodele
  Cole, Greg Grabeel, Eric Gustafson, Chris Guido, Matthew Heintze, Alastair
  Heptonstall, Mindy Jacobson, Won Kim, Eleanor King, Alexander Lynch, Stephen
  O'Connor, David Ottaway, Ken Mailand, Guido Mueller, Jesper Munch, Virginio
  Sannibale, Zhenhua Shao, Michael Smith, Peter Veitch, Thomas Vo, Cheryl
  Vorvick, and Phil Willems.
\newblock {Overview of Advanced LIGO adaptive optics}.
\newblock {\em Appl. Opt.}, 55(29):8256--8265, October 2016.

\bibitem{Yamamoto:2008}
Hiro Yamamoto.
\newblock {SIS (Stationary Interferometer Simulation) manual}.
\newblock LIGO Technical Report LIGO-T070039, 2008.

\bibitem{Pinard:2017}
L.~Pinard, C.~Michel, B.~Sassolas, L.~Balzarini, J.~Degallaix, V.~Dolique,
  R.~Flaminio, D.~Forest, M.~Granata, B.~Lagrange, N.~Straniero, J.~Teillon,
  and G.~Cagnoli.
\newblock Mirrors used in the {LIGO} interferometers for first detection of
  gravitational waves.
\newblock {\em Appl. Opt.}, 56(4):C11--C15, Feb 2017.

\bibitem{Evans:2015}
Matthew Evans, Slawek Gras, Peter Fritschel, John Miller, Lisa Barsotti, Denis
  Martynov, Aidan Brooks, Dennis Coyne, Rich Abbott, Rana~X. Adhikari, Koji
  Arai, Rolf Bork, Bill Kells, Jameson Rollins, Nicolas Smith-Lefebvre,
  Gabriele Vajente, Hiroaki Yamamoto, Carl Adams, Stuart Aston, Joseph
  Betzweiser, Valera Frolov, Adam Mullavey, Arnaud Pele, Janeen Romie, Michael
  Thomas, Keith Thorne, Sheila Dwyer, Kiwamu Izumi, Keita Kawabe, Daniel Sigg,
  Ryan Derosa, Anamaria Effler, Keiko Kokeyama, Stefan Ballmer, Thomas~J.
  Massinger, Alexa Staley, Matthew Heinze, Chris Mueller, Hartmut Grote, Robert
  Ward, Eleanor King, David Blair, Li~Ju, and Chunnong Zhao.
\newblock Observation of parametric instability in {Advanced LIGO}.
\newblock {\em Phys. Rev. Lett.}, 114:161102, Apr 2015.

\bibitem{McCuller:2021}
L.~McCuller, S.~E. Dwyer, A.~C. Green, Haocun Yu, K.~Kuns, L.~Barsotti, C.~D.
  Blair, D.~D. Brown, A.~Effler, M.~Evans, A.~Fernandez-Galiana, P.~Fritschel,
  V.~V. Frolov, N.~Kijbunchoo, G.~L. Mansell, F.~Matichard, N.~Mavalvala, D.~E.
  McClelland, T.~McRae, A.~Mullavey, D.~Sigg, B.~J.~J. Slagmolen, M.~Tse,
  T.~Vo, R.~L. Ward, C.~Whittle, R.~Abbott, C.~Adams, R.~X. Adhikari,
  A.~Ananyeva, S.~Appert, K.~Arai, J.~S. Areeda, Y.~Asali, S.~M. Aston,
  C.~Austin, A.~M. Baer, M.~Ball, S.~W. Ballmer, S.~Banagiri, D.~Barker,
  J.~Bartlett, B.~K. Berger, J.~Betzwieser, D.~Bhattacharjee, G.~Billingsley,
  S.~Biscans, R.~M. Blair, N.~Bode, P.~Booker, R.~Bork, A.~Bramley, A.~F.
  Brooks, A.~Buikema, C.~Cahillane, K.~C. Cannon, X.~Chen, A.~A. Ciobanu,
  F.~Clara, C.~M. Compton, S.~J. Cooper, K.~R. Corley, S.~T. Countryman, P.~B.
  Covas, D.~C. Coyne, L.~E.~H. Datrier, D.~Davis, C.~Di~Fronzo, K.~L. Dooley,
  J.~C. Driggers, T.~Etzel, T.~M. Evans, J.~Feicht, P.~Fulda, M.~Fyffe, J.~A.
  Giaime, K.~D. Giardina, P.~Godwin, E.~Goetz, S.~Gras, C.~Gray, R.~Gray, E.~K.
  Gustafson, R.~Gustafson, J.~Hanks, J.~Hanson, T.~Hardwick, R.~K. Hasskew,
  M.~C. Heintze, A.~F. Helmling-Cornell, N.~A. Holland, J.~D. Jones,
  S.~Kandhasamy, S.~Karki, M.~Kasprzack, K.~Kawabe, P.~J. King, J.~S. Kissel,
  Rahul Kumar, M.~Landry, B.~B. Lane, B.~Lantz, M.~Laxen, Y.~K. Lecoeuche,
  J.~Leviton, J.~Liu, M.~Lormand, A.~P. Lundgren, R.~Macas, M.~MacInnis, D.~M.
  Macleod, S.~M\'arka, Z.~M\'arka, D.~V. Martynov, K.~Mason, T.~J. Massinger,
  R.~McCarthy, S.~McCormick, J.~McIver, G.~Mendell, K.~Merfeld, E.~L. Merilh,
  F.~Meylahn, T.~Mistry, R.~Mittleman, G.~Moreno, C.~M. Mow-Lowry, S.~Mozzon,
  T.~J.~N. Nelson, P.~Nguyen, L.~K. Nuttall, J.~Oberling, Richard~J. Oram,
  C.~Osthelder, D.~J. Ottaway, H.~Overmier, J.~R. Palamos, W.~Parker, E.~Payne,
  A.~Pele, R.~Penhorwood, C.~J. Perez, M.~Pirello, H.~Radkins, K.~E. Ramirez,
  J.~W. Richardson, K.~Riles, N.~A. Robertson, J.~G. Rollins, C.~L. Romel,
  J.~H. Romie, M.~P. Ross, K.~Ryan, T.~Sadecki, E.~J. Sanchez, L.~E. Sanchez,
  T.~R. Saravanan, R.~L. Savage, D.~Schaetzl, R.~Schnabel, R.~M.~S. Schofield,
  E.~Schwartz, D.~Sellers, T.~Shaffer, J.~R. Smith, S.~Soni, B.~Sorazu, A.~P.
  Spencer, K.~A. Strain, L.~Sun, M.~J. Szczepa\ifmmode~\acute{n}\else
  \'{n}\fi{}czyk, M.~Thomas, P.~Thomas, K.~A. Thorne, K.~Toland, C.~I. Torrie,
  G.~Traylor, A.~L. Urban, G.~Vajente, G.~Valdes, D.~C. Vander-Hyde, P.~J.
  Veitch, K.~Venkateswara, G.~Venugopalan, A.~D. Viets, C.~Vorvick, M.~Wade,
  J.~Warner, B.~Weaver, R.~Weiss, B.~Willke, C.~C. Wipf, L.~Xiao, H.~Yamamoto,
  Hang Yu, L.~Zhang, M.~E. Zucker, and J.~Zweizig.
\newblock Ligo's quantum response to squeezed states.
\newblock {\em Phys. Rev. D}, 104:062006, Sep 2021.

\bibitem{Miranda:2018}
Lester~James Miranda.
\newblock Pyswarms: a research toolkit for particle swarm optimization in
  {Python}.
\newblock {\em Journal of Open Source Software}, 3(21):433, 2018.

\bibitem{FinesseRef}
Daniel~David Brown and Andreas Freise.
\newblock Finesse, May 2014.
\newblock {You can download the binaries and source code at
  \url{http://www.gwoptics.org/finesse}.}

\bibitem{Brown:2020}
Daniel~D. Brown, Philip Jones, Samuel Rowlinson, Sean Leavey, Anna~C. Green,
  Daniel Töyrä, and Andreas Freise.
\newblock Pykat: Python package for modelling precision optical
  interferometers.
\newblock {\em SoftwareX}, 12:100613, 2020.

\bibitem{Fritschel:2014}
Peter Fritschel, Matthew Evans, and Valery Frolov.
\newblock {Balanced homodyne readout for quantum limited gravitational wave
  detectors}.
\newblock {\em Opt. Express}, 22(4):4224--4234, February 2014.

\bibitem{Martynov:2016}
D.~V. Martynov, E.~D. Hall, B.~P. Abbott, R.~Abbott, T.~D. Abbott, C.~Adams,
  R.~X. Adhikari, R.~A. Anderson, S.~B. Anderson, K.~Arai, M.~A. Arain, S.~M.
  Aston, L.~Austin, S.~W. Ballmer, M.~Barbet, D.~Barker, B.~Barr, L.~Barsotti,
  J.~Bartlett, M.~A. Barton, I.~Bartos, J.~C. Batch, A.~S. Bell, I.~Belopolski,
  J.~Bergman, J.~Betzwieser, G.~Billingsley, J.~Birch, S.~Biscans, C.~Biwer,
  E.~Black, C.~D. Blair, C.~Bogan, C.~Bond, R.~Bork, D.~O. Bridges, A.~F.
  Brooks, D.~D. Brown, L.~Carbone, C.~Celerier, G.~Ciani, F.~Clara, D.~Cook,
  S.~T. Countryman, M.~J. Cowart, D.~C. Coyne, A.~Cumming, L.~Cunningham,
  M.~Damjanic, R.~Dannenberg, K.~Danzmann, C.~F. Da~Silva Costa, E.~J. Daw,
  D.~DeBra, R.~T. DeRosa, R.~DeSalvo, K.~L. Dooley, S.~Doravari, J.~C.
  Driggers, S.~E. Dwyer, A.~Effler, T.~Etzel, M.~Evans, T.~M. Evans,
  M.~Factourovich, H.~Fair, D.~Feldbaum, R.~P. Fisher, S.~Foley, M.~Frede,
  A.~Freise, P.~Fritschel, V.~V. Frolov, P.~Fulda, M.~Fyffe, V.~Galdi, J.~A.
  Giaime, K.~D. Giardina, J.~R. Gleason, R.~Goetz, S.~Gras, C.~Gray, R.~J.~S.
  Greenhalgh, H.~Grote, C.~J. Guido, K.~E. Gushwa, E.~K. Gustafson,
  R.~Gustafson, G.~Hammond, J.~Hanks, J.~Hanson, T.~Hardwick, G.~M. Harry,
  K.~Haughian, J.~Heefner, M.~C. Heintze, A.~W. Heptonstall, D.~Hoak, J.~Hough,
  A.~Ivanov, K.~Izumi, M.~Jacobson, E.~James, R.~Jones, S.~Kandhasamy,
  S.~Karki, M.~Kasprzack, S.~Kaufer, K.~Kawabe, W.~Kells, N.~Kijbunchoo, E.~J.
  King, P.~J. King, D.~L. Kinzel, J.~S. Kissel, K.~Kokeyama, W.~Z. Korth,
  G.~Kuehn, P.~Kwee, M.~Landry, B.~Lantz, A.~Le~Roux, B.~M. Levine, J.~B.
  Lewis, V.~Lhuillier, N.~A. Lockerbie, M.~Lormand, M.~J. Lubinski, A.~P.
  Lundgren, T.~MacDonald, M.~MacInnis, D.~M. Macleod, M.~Mageswaran,
  K.~Mailand, S.~M\'arka, Z.~M\'arka, A.~S. Markosyan, E.~Maros, I.~W. Martin,
  R.~M. Martin, J.~N. Marx, K.~Mason, T.~J. Massinger, F.~Matichard,
  N.~Mavalvala, R.~McCarthy, D.~E. McClelland, S.~McCormick, G.~McIntyre,
  J.~McIver, E.~L. Merilh, M.~S. Meyer, P.~M. Meyers, J.~Miller, R.~Mittleman,
  G.~Moreno, C.~L. Mueller, G.~Mueller, A.~Mullavey, J.~Munch, P.~G. Murray,
  L.~K. Nuttall, J.~Oberling, J.~O'Dell, P.~Oppermann, Richard~J. Oram,
  B.~O'Reilly, C.~Osthelder, D.~J. Ottaway, H.~Overmier, J.~R. Palamos, H.~R.
  Paris, W.~Parker, Z.~Patrick, A.~Pele, S.~Penn, M.~Phelps, M.~Pickenpack,
  V.~Pierro, I.~Pinto, J.~Poeld, M.~Principe, L.~Prokhorov, O.~Puncken,
  V.~Quetschke, E.~A. Quintero, F.~J. Raab, H.~Radkins, P.~Raffai, C.~R. Ramet,
  C.~M. Reed, S.~Reid, D.~H. Reitze, N.~A. Robertson, J.~G. Rollins, V.~J.
  Roma, J.~H. Romie, S.~Rowan, K.~Ryan, T.~Sadecki, E.~J. Sanchez, V.~Sandberg,
  V.~Sannibale, R.~L. Savage, R.~M.~S. Schofield, B.~Schultz, P.~Schwinberg,
  D.~Sellers, A.~Sevigny, D.~A. Shaddock, Z.~Shao, B.~Shapiro, P.~Shawhan,
  D.~H. Shoemaker, D.~Sigg, B.~J.~J. Slagmolen, J.~R. Smith, M.~R. Smith, N.~D.
  Smith-Lefebvre, B.~Sorazu, A.~Staley, A.~J. Stein, A.~Stochino, K.~A. Strain,
  R.~Taylor, M.~Thomas, P.~Thomas, K.~A. Thorne, E.~Thrane, K.~V. Tokmakov,
  C.~I. Torrie, G.~Traylor, G.~Vajente, G.~Valdes, A.~A. van Veggel, M.~Vargas,
  A.~Vecchio, P.~J. Veitch, K.~Venkateswara, T.~Vo, C.~Vorvick, S.~J. Waldman,
  M.~Walker, R.~L. Ward, J.~Warner, B.~Weaver, R.~Weiss, T.~Welborn,
  P.~We\ss{}els, C.~Wilkinson, P.~A. Willems, L.~Williams, B.~Willke,
  I.~Wilmut, L.~Winkelmann, C.~C. Wipf, J.~Worden, G.~Wu, H.~Yamamoto, C.~C.
  Yancey, H.~Yu, L.~Zhang, M.~E. Zucker, and J.~Zweizig.
\newblock Sensitivity of the {Advanced LIGO} detectors at the beginning of
  gravitational wave astronomy.
\newblock {\em Phys. Rev. D}, 93:112004, Jun 2016.

\bibitem{Staley:2014}
A~Staley, D~Martynov, R~Abbott, R~X Adhikari, K~Arai, S~Ballmer, L~Barsotti,
  A~F Brooks, R~T DeRosa, S~Dwyer, A~Effler, M~Evans, P~Fritschel, V~V Frolov,
  C~Gray, C~J Guido, R~Gustafson, M~Heintze, D~Hoak, K~Izumi, K~Kawabe, E~J
  King, J~S Kissel, K~Kokeyama, M~Landry, D~E McClelland, J~Miller, A~Mullavey,
  B~O'Reilly, J~G Rollins, J~R Sanders, R~M~S Schofield, D~Sigg, B~J~J
  Slagmolen, N~D Smith-Lefebvre, G~Vajente, R~L Ward, and C~Wipf.
\newblock Achieving resonance in the {Advanced LIGO} gravitational-wave
  interferometer.
\newblock {\em Classical and Quantum Gravity}, 31(24):245010, nov 2014.

\bibitem{Kogelnik:1966}
H.~Kogelnik and T.~Li.
\newblock Laser beams and resonators.
\newblock {\em Appl. Opt.}, 5(10):1550--1567, Oct 1966.

\bibitem{Arai:2013}
Koji Arai.
\newblock {On the accumulated round-trip Gouy phase shift for a general optical
  cavity}.
\newblock LIGO Technical Report LIGO-T1300189-v1, 2013.

\bibitem{Bochner:2003}
Brett {Bochner}.
\newblock Simulating a dual-recycled gravitational wave interferometer with
  realistically imperfect optics.
\newblock {\em General Relativity and Gravitation}, 35(6):1029--1057, June
  2003.

\end{thebibliography}

\end{document}